\documentclass[proof]{WileyASNA-v1}

\articletype{Article Type}%

\usepackage [latin1]{inputenc}

\received{05 april 2021}
\revised{11 april 2021}
\accepted{17 april 2021}

\begin{document}

\title{Pushing the limits of  time beyond the Big Bang singularity: Scenarios for the branch cut universe}

\author[1,2]{C\'esar A. Zen Vasconcellos*}

\author[3,4]{Peter O. Hess}

\author[1]{Dimiter Hadjimichef}

\author[5]{Benno Bodmann}

\author[6]{Mois\'es Razeira}

\author[1]{Guilherme L. Volkmer} 

\authormark{C\'esar A. Zen Vasconcellos, Peter O. Hess, Dimiter Hadjimichef, Benno Bodmann, Mois\'es Razeira, and Guilherme L. Volkmer }

\address[1]{\orgdiv{Instituto de F\'isica}, \orgname{Universidade Federal do Rio Grande do Sul (UFRGS)}, \orgaddress{\state{Porto Alegre}, \country{Brazil}}}

\address[2]{\orgdiv{International Center for Relativistic Astrophysics Network (ICRANet), Pescara, Italy}}  

\address[3]{Universidad Nacional Aut\'onoma de Mexico (UNAM), M\'exico City, M\'exico}

\address[4]{Frankfurt Institute for Advanced Studies (FIAS), J.W. von Goethe University (JWGU), Hessen, Germany}

\address[5]{\orgdiv{Unversidade Federal de Santa Maria (UFSM), Santa Maria, Brazil}}

\address[6]{\orgdiv{Laborat\'orio de Geoci\^encias Espaciais e Astrof\'isica (LaGEA)}, \orgname{Universidade Federal do PAMPA (UNIPAMPA)}, \orgaddress{\state{Ca\c{c}apava do Sul}, \country{Brazil}}}

\corres{*Av. Bento Gonçalves, 9500 - Agronomia, Porto Alegre - RS, 91501-970.  \email{cesarzen@cesarzen.com}}

\abstract{In this contribution we identify two scenarios for the evolutionary branch cut universe. In the first scenario, the universe evolves continuously from the negative complex cosmological time sector, prior to a primordial singularity, to the positive one, circumventing continuously a branch cut, and no primordial singularity occurs in the imaginary sector, only branch points. In the second scenario, the branch cut and branch point disappear after the {\it realisation} of the imaginary component of the complex time by means of a Wick rotation, which is replaced by the thermal time. In the second scenario, the universe has its origin in the Big Bang, but the model contemplates simultaneously a mirrored parallel evolutionary universe going backwards in the cosmological thermal time negative sector. A quantum formulation based on the WDW equation is sketched and preliminary conclusions are drawn.} 


\maketitle

\section{The illusion of time}

Newton's conception of the {\it clockwork universe}, evolving like a mechanical perfect clock and whose movements of its gears are governed by the laws of physics, with an inherent predictability, prevailed for more than three centuries, until a revolutionary concept emerged, thanks to the genius mind of 
Hermann Minkowski\citep{Minkowski1915}, with profound consequences for our current understanding of its structure and evolution. 

According to his view, instead of being considered separate entities (though intimately related), space and time were combined into a single continuum entity, the spacetime\footnote{In Minkowski's own words: ``Henceforth space by itself and time by itself are doomed to fade away into mere shadows, and only a kind of union of the two will preserve an independent
reality''~\citep{Minkowski1915}.}.   

Time in physics is usually considered a fundamental variable,
defined by its measurement as the reading of a clock. In opposition to this view, 
John Weeler and Bryce DeWitt developed in 1967 the so called WdW equation~\citep{WdW}  based on the audacious idea of {\it physics without time}, a theoretical framework that sought to combine quantum mechanics and general relativity, representing a step towards a consistent theory of quantum gravity.

More recently, Carlo Rovelli affirmed that the flow of time is an illusion and that our naive perception of it doesn't correspond to physical reality~\citep{Rovelli2019}, a vision that is in tune with Albert Einstein's perception of time\footnote{Albert Einstein in a letter to the family of Michele Besso, his collaborator and closest friend, once wrote:
``Now he has departed from this strange world a little ahead of me. That means nothing. People like us, who believe in physics, know that the distinction between past, present, and future is only a stubbornly persistent illusion.''~\citep{CHRISTIES2020}.
}. Carlos Rovelli also recently revisited the idea of `physics without time'~\citep{Rovelli2004,Rovelli2011,Rovelli2015} bearing in mind that, in accordance with the second law of thermodynamics,  {\it forward in time} is the direction in which entropy increases, and in which we gain information, so the flow of time is a subjective feature of the universe, not an objective part of physical reality\footnote{In general relativity, the {\it reading of a clock}  is not given by the time variable $t$, but is instead expressed by a line integral depending on the gravitational field, computed along the clock's world-line $\gamma$, given as
$$
{\cal{T}} = \int_{\gamma} \sqrt{g_{\mu\nu}(x,t) \, dx^{\mu}dx^{\nu}} \, ;
$$
(see~\citet{Rovelli2015} for more details).}. 
In this realm,  in which the observable universe does not show  time reversal symmetry, 
events, rather than particles or fields, are the basic constituents of the universe, and the task of physics would be to describe the relationship between events. 

These conceptions present some points of contact  with a line of thought we recently developed~\citep{Zen2020,ZenA} where we applied  the tools of singular semi-Riemannian geometry to push the limits of general relativity and time beyond a primordial singularity, giving rise to a branch cut universe. In this contribution we sought to identify evolutionary scenarios for the branch cut universe. 

\section{Wick rotation of cosmological time}

Wick rotation is a well known theoretical method that encapsulates a connection  between quantum mechanics and quantum statistical mechanics and in another ground relativistic field theory in Minkowski spacetime manifolds and Euclidean field theory in Riemannian spacetime manifolds.
 
 \subsection{Path integral formalism}
 
 The path integral formalism~\citep{Feynman1965}
 describes the quantum transition amplitudes of the unitary time evolution operator,
 $ \hat{U}(t_{\rm{b}}, t_{\rm{a}})$ (a representation of the abelian group
of time translations), 
between the localised quantum mechanical states of a particle
($x_a, t_{\rm{a}}$) to ($x_b, t_{\rm{b}}$), with $x$ and $t$ denoting space and time Cartesian coordinates.
The matrix
elements of the quantum time evolution amplitudes, using 
bra's ($\langle x_b |$) and ket's ($ |x_a \rangle$)
notation,
read
\begin{equation}
(x_bt_{\rm{b}}|x_at_{\rm{a}}) = \langle x_b |  \hat{U}(t_{\rm{b}}, t_{\rm{a}}) |x_a \rangle \, \, t_{\rm{b}} > t_{\rm{a}} \, ,  \label{EA}
 \end{equation} 
 
For a system with a time-independent
Hamiltonian operator, $\hat{H}$, 
 the time evolution operator is simply
\begin{equation}
 \hat{U}(t_{\rm{b}}, t_{\rm{a}}) = \hat{\bar{\tau}} e^{-\frac{i}{\hbar}\hat{H} (t_{\rm{b}} - t_{\rm{a}})} \, , 
 \end{equation}
where  $\hat{\bar{\tau}}$  denotes the time-ordering operator. In the continuum limit, we write the amplitude $(x_bt_{\rm{b}}|x_at_{\rm{a}})$ as a path integral
\begin{equation}
(x_bt_{\rm{b}}|x_at_{\rm{a}}) \equiv \int_{x_a}^{x_b} {\cal D}_x e^{i {\cal A}(x)/\hbar} \, .
\end{equation}
This equation is the corresponding Feynman's formula for the quan\-tum-me\-cha\-ni\-cal amplitude (\ref{EA}) and represents the
sum over all paths in configuration space with a phase factor containing
the action ${\cal A}[x]$. 

 \subsection{Wick rotation in statistical and quantum mechanics}
 
In statistical mechanics, the quantum partition function $Z(T)$,
 which
contains all information about the thermodynamical equilibrium properties of a quantum
system,
reads
\begin{equation}
Z(T) \equiv Tr \Bigl( e^{- \hat{H}/k_B T}  \Bigr)  \equiv Tr \Bigl( e^{- H(\hat{p}, \hat{x}) /k_B T} \Bigr) \, .  
\end{equation}
In this expression, $Tr(\hat{F})$ denotes the trace of the operator 
\begin{equation}
\hat{F} = e^{- H(\hat{p}, \hat{x}) /k_B T} \, , 
\end{equation}
and $k_B$ is the Boltzmann constant.  For a $N$-particle system described by the Schr\"odinger equation for instance, the quantum-statistical system refers to a canonical ensemble.  

The quantum statistical partition function $Z(T)$
may be related to the quantum-mechanical time evolution operator $Z_{QM}(t_{\rm{b}} - t_{\rm{a}})$ 
\begin{equation}
Z_{QM}(t_{\rm{b}} - t_{\rm{a}}) \equiv Tr \Bigl[ \hat{U}(t_{\rm{b}},t_{\rm{a}}) \Bigr] = Tr \Bigl[ e^{-i(t_{\rm{b}} - t_{\rm{a}})\hat{H}/\hbar}  \Bigr] 
\, ,  \label{WR}
\end{equation}
by making an analytical continuation  of the time interval $t_{\rm{b}}- t_{\rm{a}}$ to the negative imaginary value using a Wick rotation:
\begin{equation}
t_{\rm{b}} - t_{\rm{a}} \rightarrow   - i \hbar / k_BT \, .
\end{equation}
In quantum mechanics and quantum field theory, the Hamiltonian density ${\cal H}$ acts as the generator of the Lie group of time translations while in statistical mechanics the Hamiltonian represents a Boltzmann weight in an ensemble. The contour-rotation from the real to the imaginary time-axis, results in a correspondence between the imaginary time component and the inverse of the temperature, $T$ and it can be understood as a realisation of the imaginary component of time.

 \subsection{Euclidean quantum gravity}

Euclidean quantum gravity refers to a quantum theory of Riemannian manifolds in which the quantisation of gravity occurs in a Euclidean spacetime, generated by means of a Wick rotation. The corresponding gravitational path integral in the presence of a field $\phi$ may be expressed as
\begin{equation}
{\cal Z} = \int {\cal D}[g] {\cal D}[\phi] e^{\int d^4x \sqrt{|{\mathbf g}| } \,R} \, . 
\end{equation}
Additional assumptions imposed to the manifolds as compactness, connectivity and boundaryless (no singularities), make this formulation a strong candidate for overcoming the limitations presented by general relativity in the domain of strong gravity, more precisely, the elimination of singularities in extreme physical conditions.  There are other techniques that sought to overcome these limitations of general relativity. Among these, we highlight the pseudo-complex general relativity, a very powerful technique based on pseudo-complex spacetime coordinates~\citep{PNP2020,Hess2017,Hess2020} with observational predictions given in \citep{MNRAS2014}.

\section{Cosmography in an universe with a branch cut}

The tracking of the analytically continued scale factor $\ln^{-1}[\beta(t)]$ and 
the background cosmological Hubble rate $H_{\rm{ac}}(t)$, analytically continued to the complex plane
enable us to trace the evolutive paths  of the branch cut universe from its initial stages to the present days (for the details see~\citet{ZenA}). 

The scale factor $\ln^{-1}[\beta(t)]$, a dimensionless quantity, describes the change in sizes of {\it portions of space} (or patches) due to the expansion or contraction of the branch cut universe. The Hubble parameter  $H_{\rm{ac}}(t)$ in turn measures the expansion rate of the branch cut universe. We assume the observable universe corresponds today to a patch of space with radius $R(t_0)$, and that the patch size of the branch cut universe at any other period of time is given by 
\begin{equation}\frac{\ln^{-1}[\beta(t)]}{\ln^{-1}(\beta_0)} R(t_0) = 
\frac{\ln(\beta_0)}{\ln[\beta(t)]} R(t_0) 
\, . \end{equation} 

\subsection{Cosmological parameters}

The analytically continued energy-stress conservation law in the expanding universe may be written as (for the details see~\citet{ZenA}):
\begin{eqnarray}
 \frac{1}{\rho(t)}\frac{d}{dt}\rho(t)   \! + \!   3 \Bigl(\! 1 + \frac{p(t)}{c^2\rho(t)}  \! \Bigr) \frac{1}{\ln^{-1}[\beta(t)]} \frac{d\ln^{-1}[\beta(t)]}{dt} \label{CE}
 \\
 \!  \Rightarrow \!  \frac{d}{dt}\ln  (\rho(t)) \! + \! 
3 \Bigl(1 + \frac{p(t)}{c^2\rho(t)}  \Bigr)  \frac{d}{dt} \ln [\ln^{-1}[\beta(t)]  \! = \! 0.  \nonumber \end{eqnarray}
From this equation it results 
\begin{equation}
\rho(t) = \rho_0 exp \Biggl( - 2 \int \epsilon(t)  d \!\! \ln \, (\ln^{-1}[\beta(t)])   \Biggr) \, , \label{ro}
\end{equation}
where
\begin{equation}
 \epsilon(t) \equiv \frac{3}{2} \Bigl(1 + \frac{p(t)}{c^2 \rho(t)}  \Bigr) \, , \label{e}
\end{equation}
represents a dimensionless thermodynamical connection between the energy density $\rho(t)$ and pressure $p(t)$ of a perfect fluid thus enabling the fully description of the equation of state (EoS) of the system. Positive pressure corresponds to $\epsilon > 3/2$,  negative pressure to  $\epsilon < 3/2$ and for a universe
dominated by a cosmological constant, $\epsilon \to 0$. 

In the limit in which the 
dimensionless thermodynamical connection obeys 
$\epsilon(t) \to  \epsilon = $ constant, the integral (\ref{ro}) reduces to
\begin{eqnarray}
\ln[\rho(t)/\rho_0] & = & - 2  \lim_{\epsilon(t) \to \epsilon} \int \epsilon(t)  \,d\!\! \ln (\ln^{-1}[\beta(t)])\nonumber 
\\ & \simeq& -  2 \epsilon \ln (\ln^{-1}[\beta(t)])  \Rightarrow \ln \, (\ln^{-1}[\beta(t)])^{-2\epsilon} \nonumber \\
 & \Rightarrow  & \rho(t) \simeq \frac{\rho_0}{\ln^{-2\epsilon}[\beta(t)]} \, , \label{dens}
\end{eqnarray}
which corresponds to an analytically continued expression for the density of the branch cut universe.
Applying complex conjugation to this expression we get
\begin{equation}
 \rho^{*}(t^{*}) = \frac{\rho^*_0}{\ln^{-2\epsilon}(\beta^{*}(t^{*}))} \, .
\end{equation}

\subsection{Horizons and cosmological curvatures}

An event's causality is limited to its frontal light cone since information cannot travel faster than the speed of light. Light rays travel in null geodesics, so the following  expressions results for the analytically continued (\rm{ac})
 co-moving (\rm{cm}), ${\cal D}^{\rm{cm}}_{\rm{ac}}(t)$,  and proper (\rm{p}), ${\cal D}^{\rm{p}}_{\rm{ac}}(t)$, distances  to the horizon: \begin{equation}
{\cal D}^{\rm{cm}}_{\rm{ac}}(t) =  \int_{t_{\rm{P}}}^{t} \frac{c dt}{\ln^{-1}[\beta(t)]} \, ; \,\,\,
{\cal D}^{\rm{p}}_{\rm{ac}}(t) =  \ln^{-1}[\beta(t)] \int_{t_{\rm{P}}}^{t} \frac{c dt^{\prime}}{\ln^{-1}[\beta(t^{\prime})]} \, . 
\end{equation}
We also develop expressions for the analytically continued time-dependent and dimensionless cosmic curvature factor (\rm{ccf}), $\Omega^{\rm{ccf}}_{\rm{ac}}(t)$ (apparent spatial curvature), and the cosmic anisotropy factor (\rm{caf}), $\Omega^{\rm{caf}}_{\rm{ac}}(t)$ (apparent anisotropy):
\begin{equation}
\Omega^{\rm{ccf}}_{\rm{ac}}(t) \! = \! - \frac{kc^2}{\ln^{-2}[\beta(t)]} H_{\rm{ac}}^{-2}(t)  \, ; \,\,\,
\Omega^{\rm{caf}}_{\rm{ac}}(t)  \! = \!  \frac{\sigma^2}{\ln^{-6}[\beta(t)]} H_{\rm{ac}}^{-2}(t) \, .
\end{equation}
Combining these expressions with the definition of $H_{\rm{ac}}$~\citep{ZenA}, we get
\begin{equation}
\Omega^{\rm{ccf}}_{\rm{ac}}(t)  =  - \frac{k\beta^2(t)}{\dot{\beta}^2(t)\ln^{-4}[\beta(t)]} \,  ; \,\,\,\Omega^{\rm{caf}}_{\rm{ac}}(t)  =  \frac{\sigma^2\beta^2(t)}{\dot{\beta}^2(t)\ln^{-8}[\beta(t)]}. 
\end{equation}
Applying complex conjugation to these expressions we obtain  
\begin{equation}
\Omega^{*\rm{ccf}}_{\rm{ac}}(t^*)  =  - \frac{k\beta^{*2}(t^*)}{\dot{\beta}^{*2}(t^*)\ln^{-4}[\beta^*(t^*)]} \, , 
\end{equation}
\begin{equation}
\mbox{and} \,\,\,\, \Omega^{*\rm{caf}}_{\rm{ac}}(t^*)  = \frac{\sigma^{*2}\beta^{*2}(t^*)}{\dot{\beta}^{*2}(t^*)\ln^{-8}[\beta^*(t^*)]}  \, .
\end{equation}
In  appendix A, we present solutions for the cosmography parameters in a branch cut universe.

%%%%%%%%%%%%%%%%%
\subsection{Cosmological Redshift}
%%%%%%%%%%%%%%%%%%

Light emitted by distant objects from our galaxy travels from the point of emission
at $t = t_{\rm{e}}$, $r = r_{\rm{e}}$
 to the observation point today  
$t = t_{\rm{o}}$, $r = r_{\rm{o}}$
along the geodesic curves of a manifold, which correspond essentially to local
straight lines ($d\theta \sim d\phi \sim 0$), satisfying $ds_\xi^2 = 0$. 

The line element of the modified FLWR metric 
of a three-dimensional spatial {\it slice} of an analytically continued spacetime, in co-moving
coordinates may be written as 
\begin{equation}
ds_\xi^{2}  =  c^2 dt_\xi^{2}  -  a_\xi^{2}(t)\left(\frac{dr_\xi^{2}}{1-k\,r_\xi^{2}}+r_\xi^{2}\Bigl[d\theta^2+\sin^2\theta\,d\phi^2 \Bigr]\right) . \label{zac}
\end{equation}
The conditions $d\theta \sim d\phi \sim 0$ and
$ds_\xi^2 = 0$ applied to this equation allows the mapping\footnote{Caution should be taken here on the mapping $a_{\xi}(t) \to \ln(\beta(t))$. This mapping {\it is not} a simple direct parametrization of the scale factor based on the real FLRW single-pole metric. 
Or the result of a direct generalization of Friedmann's equations.
 Due to the non-linearity of Einstein's equations, such a direct generalisation would not be formally consistent. The present formulation 
is the outcome of complexifying the FLRW metric and results in a sum of equations associated to infinitely many poles (in tune with Hawking's assumption of infinite number of primordial universes that occurred simultaneously) arranged along a line in the complex plane with infinitesimal residues (for the details see~\citet{Zen2020}). The multiverse conception corresponds in our formulation to a theoretical mathematical device for implementation of the proposal.
}~\citep{Zen2020,ZenA} 
\begin{eqnarray}
& \rightarrow &  c^2 dt_\xi^2-  a_\xi^2(t) \frac{dr_\xi^2}{1-k\,r_\xi^2}  = 0 
 \stackrel{}{\longrightarrow}   \frac{1}{1 + z_{\rm{ac}}}  \equiv  \frac{\ln^{-1}[\beta(t)]}{\ln^{-1}(\beta_0)}, \nonumber \\
& \longrightarrow &  z_{\rm{ac}}  \equiv  \frac{\ln[\beta(t)] - \ln(\beta_0)}{\ln(\beta_0)} \, ; \label{zac2}
\end{eqnarray}
$z_{\rm{ac}}$ represents the analytically continued cosmological corresponding to the metric  (\ref{zac}), $t$ denotes the proper time measured by a co-moving observer and  the radial and angular coordinates in the co-moving frame are represented by $r$, $\theta$ and $\phi$.
From equation (\ref{zac2}), variations of $z_{\rm{ac}}$, more specifically $\Delta z$ obey 
\begin{equation}
\Delta z_{\rm{ac}} =  \frac{\ln[\beta(t)]/\beta_0}{ln(\beta_0)} \, . 
\end{equation}

A Taylor expansion of the analytically continued Hubble's law for two objects at a distance $d$ apart gives:
\begin{eqnarray}
\ln^{-1}[\beta(t)]   & =  &   \ln^{-1}(\beta_0)  +   \frac{d}{dt} \Bigl( \ln^{-1}[\beta(t)] \Bigr) \Bigr|_{t = t_0} \Bigl( t - t_0 \Bigr) \nonumber \\
 & + &  \frac{1}{2} \frac{d^2}{dt^2} \Bigl( \ln^{-1}[\beta(t)] \Bigr) \Bigr|_{t = t_0} \Bigl( t - t_0 \Bigr)^2 + \cdot\cdot\cdot 
\end{eqnarray} 
On small scales, the distance to an emitter, $d$ is approximately related to the time of emission, $t$, so 
we can then rewrite (\ref{zac2}) as
\begin{equation}
 \frac{1}{1 + z_{\rm{ac}}}  - H_{\rm{ac}0} \frac{d}{c} - \frac{q_{\rm{ac}0}}{2} H^2_{\rm{ac}0} \Bigl(\frac{d}{c}\Bigr)^2 + \cdot\cdot\cdot
\end{equation} 
with $\ln^{-1}(\beta_0)$ normalised to $1$ and where
\begin{equation}
q_{\rm{ac}0} = - \frac{\Bigl(\frac{d^2}{dt^2}\ln^{-1}[\beta(t)]\Bigr)\ln^{-1}[\beta(t)]}{\bigl(\frac{d}{dt}\ln^{-1}[\beta(t)]\bigr)^2} \, , 
\end{equation}
represents the analytically continued deceleration parameter. On small scales and at small
redshifts we obtain the analytically continued Hubble's law, \begin{equation} cz_{\rm{ac}} = H_{\rm{ac}0} d \, . \end{equation}

\section{On the road of a quantum approach}\label{QA}

The challenge of building a quantum theory of gravitation based on the simple combination of quantum mechanics and general relativity, due to their so distinct characteristics, is significant. In the following we present a few remarks about the physical and geometric meaning of $\ln^{-1}[\beta(t)]$ and $\beta(t)$ and we sketch first steps on the road of a quantum approach for the branch cut universe. 

\subsection{The problem of time}

It is believed that general relativity and quantum mechanics should be reconciled in a theory of quantum gravity, merging at the Planck scale. It is also believed that the spacetime geometry cannot be measured below the Planck scale~\citep{Garay1999,Calmet}, since quantum spacetime fluctuations would spoil at this scale its description as a smooth spacetime manifold~\citep{Garay1999}.

The essential difference between quantum mechanics and general relativity that makes their reconciliation
extremely difficult is the interpretation and the role of time.
In quantum mechanics, time corresponds to a universal and absolute parameter. As a consequence, the formal treatment of time in quantum mechanics differs from the other coordinates that  may be raised to the category of quantum operators and observables.
In general relativity, in turn,  
time is said to be malleable and relative~\citep{Isham1993}. As a consequence, unification of quantum mechanics and general relativity requires reconciling their absolute and relative notions of time. 

\subsection{Wheeler-DeWitt Equation}

An illuminating example is the formulation and interpretation of
the Wheeler-DeWitt (WdW) equation~\citep{WdW}.
The Wheeler-DeWitt formulation for quantum gravity consists in constraining a wave function which applies to the universe as a whole, --- the so called wave function of the universe ---, in accordance with the Dirac recipe: 
\begin{equation}
\hat{\cal H} \Psi = 0 \, ,
\end{equation}
i.e., a stationary, timeless equation, instead of a time-dependent quantum mechanics
wave equation as for instance
\begin{equation}
i \frac{\partial}{\partial t} \hat{H} \Psi \, . 
\end{equation}
Here, $\hat{H}$ denotes the Hamiltonian operator of a quantum subsystem,  
while $\hat{\cal H}$ in the previous equation represents a quantum operator which describes a general relativity constraining, resulting in a second order hyperbolic  equation of gravity variables\footnote{More precisely, the scale factor $a(t)$, the density $\rho(t)$, the pressure $p(t)$, and the gravitation constant $\Lambda$.}, a Klein-Gordon-type equation, having therefore a `natural' conserved  associated current (${\cal J}$),
\begin{equation}
{\cal J} = \frac{i}{2} \Bigl(\Psi^{\dagger} \, \nabla \cdot \Psi -    \Psi \,  \nabla \cdot \Psi^{\dagger} \Bigr); \, \, \, \mbox{with} \, \, \nabla \cdot {\cal J} = 0 \, .   
\end{equation}

Primordial technical deficiencies have historically led to a tendency to underestimate the WdW equation as a consistent formulation of quantum gravity, despite supporting several approaches, from quantum geometrodynamics to loop quantum gravity. More recently, however, an opposite trend has emerged, related to the understanding of the fundamental reasons for the intriguing explicit absence of the time variable in the WdW equation,
i.e., that the time variable has no physical significance in general relativity~\citep{Rovelli2011,Rovelli2015,Shestakova2018}. Based on this understanding, in the following we sketch a quantum formulation of the present approach based on the WdW equation analytically continued to the complex plane.
 
 \subsection{Analytically continued WdW equation}
 
 \subsubsection{Einstein-Hilbert action in the new metric}
 
The define the
 analytically continued FLRW metric~\citep{ZenA}:
\begin{equation}
ds_{[\rm{ac}]}^2     =    -  \sigma^2   N^2(t)  c^2dt^2   +    \sigma^2  \ln^{-2}[\beta(t)] d\Omega^2(r,\theta,\phi) \, ,  \label{FLRWac2}  
\end{equation}
with
\begin{equation}
d\Omega^2(r,\theta,\phi) \equiv
 \Biggl[ 
\frac{dr^2}{\bigl(1 -  kr^2(t) \bigr)}
  +   r^2(t) \Bigl(d \theta^2   +   sin^2 \theta d\phi^2  \Bigr)  \Biggr].   
\end{equation}
 In expression (\ref{FLRWac2}), $N(t)$ is an arbitrary lapse function\footnote{The lapse
function $N(t)$ is not dynamical, but a pure gauge variable. Gauge invariance of the action in general relativity yields a
Hamiltonian constraint which requires a gauge fixing condition on the lapse (see~\citet{Feinberg}. \label{N}}, and 
$\sigma^2 = 2/3\pi$ is just a normalisation factor.  
 
We assume as a starting point a homogeneous and isotropic multiverse described by
a mini-superspace model (see for instance~\citet{Kim}) with only one dynamical variable, the scale factor $\ln^{-1}[\beta(t)]$, and a very simple scenario for the
Einstein-Hilbert action $S_{\rm{EH}}$~\citep{He}  
 \begin{eqnarray}
 S_{\rm{EH}}  & = &  \frac{1}{16 \pi G} \int {\cal L} \, dt d^3 x   \label{S} \\
&  = &   \frac{1}{16 \pi G} \int \sqrt{-g} \Bigl(R_{[gac]} c^3- \frac{16 \pi G \rho}{\sigma^2 c}  \Bigr) dt d^3x \, ,  
  \end{eqnarray}
where $R_{[gac]}$ represents the analytically continued scalar curvature (for the details see~\citet{ZenA}):
\begin{eqnarray}
 R_{[gac]}   & = &  6 \Biggl[ \Biggl(\frac{\frac{d^2}{dt^2}{\ln^{-1}[\beta(t)]}}{\sigma^2 c^2 N^2(t) \ln^{-1}[\beta(t)]} \Biggr)  \nonumber \\
& + &  \Biggl(  \frac{\frac{d}{dt}{\ln^{-1}[\beta(t)]}}{\sigma c N(t)\ln^{-1}[\beta(t)]}   \Biggr)^2 
  +  \frac{k}{\sigma^2\ln^{-2}[\beta(t)]}  \Biggr], \label{RicciScalarac2}
\end{eqnarray}
 and $\rho$
represents the energy density of the multiverse.

Combining (\ref{S}) and (\ref{RicciScalarac2}) and using the approximation $\sqrt{-g} \approx N(t) \ln^{-3}[\beta(t)]$,  we obtain
\begin{eqnarray}
S  & \! = \! &  \int \frac{6\sigma^2N(t)c^4}{16\pi G} \Biggl(\frac{\ln^{-2}[\beta(t)]}{N^2(t) c^2} \frac{d^2}{dt^2} \ln^{-1}[\beta(t)]  \\
& \! + \! &   \ln^{-1}[\beta(t)] \Biggl[\! \Bigl(\! \frac{\frac{d}{dt} \ln^{-1}[\beta(t)]}{N(t)c} \! \Bigr)^2 + k \! \Biggr]  \! - \! \frac{8 \pi G \rho}{3c^4} \ln^{-3}[\beta(t)] \! \Biggr) \, d^4x . \nonumber
\end{eqnarray}
We integrate this equation by parts
to remove the second derivative in $\ln^{-1}[\beta(t)]$,
resulting in
\begin{eqnarray}
S  =  \frac{Nc^4}{2G} \int \Biggl(\!\!\!\! & - &  \ln^{-1}[\beta(t)]  \Bigl( \frac{\frac{d}{dt} \ln^{-1}[\beta(t)]}{Nc} \Bigr)^2  \\
& + &  k \ln^{-1}[\beta(t)]
 -  \frac{8 \pi G \rho}{3c^4} \ln^{-3}[\beta(t)] \! \Biggr) dt \, . \nonumber
\end{eqnarray}
Making the choice $N = G = c = 1$ (see footnote \ref{N}), from this equation we obtain the Lagrangian density of the multiverse:
\begin{eqnarray}
 {\cal L}  =   \frac{1}{2} \Biggl( k \ln^{-1}[\beta(t)]  \, & - & \, \ln^{-1}[\beta(t)]  \Bigl( \frac{d}{dt} \ln^{-1}[\beta(t)]  \Bigr)^2 \nonumber \\
& - &  \, \frac{8 \pi \rho}{3} \ln^{-3}[\beta(t)]  \Biggr) . 
\end{eqnarray}

In the following, on basis of this Lagrangian formulation, we proceed with the quantisation of the system.

\subsection{Topological Quantisation}\label{QF}

The conjugate momentum $p_{\rm{\ln}}$ of the dynamical variable $\ln^{-1}[\beta(t)]$ is 
\begin{equation}
p_{\rm{\ln}}  =  \frac{\partial {\cal L}}{\partial \bigl(\frac{d}{dt} \ln^{-1}[\beta(t)]\bigr)} =  - \ln^{-1}[\beta(t)] \frac{\dot{\beta}(t)}{\beta(t)} \,. 
\end{equation}

Therefore the corresponding Hamiltonian becomes
\begin{eqnarray}
{\cal H} & = & p_{\rm{\ln}} \frac{d}{dt} \ln^{-1}[\beta(t)] - {\cal L} \, , \label{H} \\
& = & - \frac{1}{2} \Biggl(\frac{p^2_{\ln}}{\ln^{-1}[\beta(t)]}  + k \ln^{-1}[\beta(t)] + \frac{8 \pi \rho}{3} \ln^{-3}[\beta(t)] \Biggr) \, .   \nonumber
\end{eqnarray} 

The quantisation of the Lagrangian density is achieved by raising the dynamical variable $\ln^{-1}[\beta(t)]$ and the conjugate momentum $p_{\rm{\ln}}$ to the category of operators   
in the form
\begin{eqnarray}
&& \ln^{-1}[\beta(t)] \rightarrow \hat{\ln}^{-1}[\beta(t)] \,  \\
&& \mbox{and} \, \, 
p_{\rm{\ln}} \rightarrow \widehat{p}_{\rm{\ln}} = - i\hbar \frac{\partial}{\partial \ln^{-1}[\beta(t)]} ; \nonumber
\end{eqnarray}
(for simplicity, in the following we skip using the hat symbol in the operators $\hat{\rm{p}}$ and  $\hat{\ln}$).
Ambiguities in ordering of the operators may be overcome by the introduction of an ordering-factor
$\alpha$ in the form
\begin{equation}
p^2 = - \frac{1}{\ln^{-\alpha}} \frac{\partial}{\partial \ln^{-1}[\beta(t)]} \Biggl( \ln^{-\alpha}[\beta(t)] \frac{\partial}{\partial \ln^{-1}[\beta(t)]} \Biggr) \, , \label{p2}
\end{equation}
with $\alpha$ usually chosen in general as $\alpha = [0,1]$;  $\alpha = 0$ corresponds to the semiclassical value; intermediate values have no meaning.  

Combining (\ref{H}) and (\ref{p2}),  using the prescription $\alpha = 0$, recovering the original values of the physical constants $ G $ and $ c $, and changing variable ($u \equiv \ln^{-1}[\beta(t)]$ and $du \equiv d\ln^{-1}[\beta(t)]$) we obtain the following expression for the WdW equation:
\begin{equation}
\Biggl( - \frac{\hbar^2}{2 m_{\rm{P}}} \frac{\partial^2}{\partial u^2}  +  \frac{E_{\rm{P}} k}{2 \ell^2_{\rm{P}}} u^2  -  \frac{4 \pi \rho}{3 \ell_{\!\!P}} u^4 \Biggr) \Psi(u)  =  0 \, ,\label{WdWSE}
\end{equation}
where $m_{\rm{P}}$, $E_{\rm{P}}$, and $\ell_{\rm{P}}$ are the Planck mass, energy and length, respectively (for comparison see for instance~\citet{He}).
This expression represents a Schr\"odinger-type equation of a particle with  the Planck mass $m_{\rm{P}}$ under the action of the  WdW quantum potential

\begin{equation}
{\cal V}_{[\rm{ac}]\rm{WdW}} \bigl(\ln^{-1}[\beta(t)]\bigr) =   \frac{E_{\rm{P}} k}{2 \ell^2_{\rm{P}}} \ln^{-2}[\beta(t)]  -  \frac{4 \pi \rho}{3 \ell_{\!\!P}} \ln^{-4}[\beta(t)]) \, .
\end{equation}

\subsubsection{Topological quantisation of spacetime}

The scale factor $\ln^{-1}[\beta (t)]$ as the only dynamical variable of the model may be raised, in a quantum approach, at the level of a quantum operator. This new status gives the scale factor $\ln^{-1}[\beta (t)] $ an additional role in describing the evolutionary process of the branch cut universe, that of representing the formal confluence between the classic description and its quantum version through a process that we call {\it spacetime topological quantisation}, as far as we know, a new nomenclature which characterises a new perspective in the quantisation of spacetime. Our formulation describes in short the relative evolution of the variable $\ln^{-1}[\beta(t)]$ over worldlines $\gamma_{\rm{\ln}} $ associated with hypersurfaces ${\cal H}_{\rm{\ln}}$ analytically continued to the complex plane. 

\subsection{Analytically continued WdW equation }

In the following we consider an extended version of the ana\-ly\-ti\-ca\-lly continued WdW equation (\ref{WdWSE}) with
extrinsic curvature contributions implemented by using the projectable Ho\v{r}ava-Lifshitz gravity model\footnote{The Ho\v{r}ava-Lifshitz formulation of gravity is an alternative theory to general relativity which
employs higher spatial-derivative terms of the curvature which are added to the Einstein-
Hilbert action with the aim of obtaining a renormalisable theory.}\footnote{For simplicity, in the following we use natural units.}:
\begin{equation}
\Bigl( - \frac{\partial^2}{\partial u^2}  +  {\cal V}_{\rm{WdWHL}}(u)  \Bigr) \Psi(u)  =  0 \, ,\label{WdWHL}
\end{equation}
with 
\begin{equation}
 {\cal V}_{\rm{WdWHL}}(u) =  \Bigl( 2  {\cal V}_{[\rm{ac}]\rm{WdW}}(u) + {\cal V}_{\rm{HL}}(u)\Bigr) \, , \label{VWDWHL}
 \end{equation}
where
${\cal V}_{\rm{HL}}(u)$ represents an adaptation of the Ho\v{r}ava-Lifshitz gravity potential:
 \begin{equation}
 {\cal V}_{\rm{HL}}(u) =  g_c k   -    g_{\Lambda} u - \frac{g_r k^2}{u} -  \frac{g_s k}{u^2} \, ; \label{VHL}
\end{equation}
(for comparison see~\citet{Orfeu,Cordero,He}).
Here, $g_C > 0$, stands for the curvature coupling constant with the sign of $g$ following the
sign of the cosmological constant~\citep{Orfeu}; $g_r$ corresponds to the coupling constant for the radiation contribution and 
$g_s$ stands for the ``stiff'' matter contribution (which corresponds to the $\rho = p$ equation of state);
$g_r$ and $g_s$ can be either positive
or negative since their signal does not alter the stability of the 
Ho\v{r}ava-Lifshitz gravity~\citep{Orfeu}. 

Combining (\ref{WdWSE}), (\ref{WdWHL}), (\ref{VWDWHL}), and (\ref{VHL}), the following equation 
results\footnote{The parameters, $g_c$, $g_{\Lambda}$, $g_r$, and $g_s$, although apparently simple, represent trick terms containing different orders of spacetime curvatures~\citep{Orfeu,Cordero,He}.}:
\begin{equation}
\Biggl( \!\! - \frac{\partial^2}{\partial u^2}  + g_c k -   g_{\Lambda}  u   + k u^2  
  -    \frac{8 \pi \rho}{3 } u^4 
 -   \frac{g_r  k^2}{u} -  \frac{g_s k}{u^2} \!\! \Biggr) \Psi(u) \! =  \! 0. \label{QWdW}
\end{equation}
This equation, when expressed in terms of non-composed functions,  as in the conventional quantum FLRW approach, is highly non-linear and has no exact solution. 

\subsubsection{Solutions}

Assuming the first two terms of the potential given by (\ref{QWdW}), $g_c k$ and  $- g_{\Lambda} u$, are dominant, the substitution
\begin{equation}
\xi \Rightarrow (g_{\Lambda})^{-2/3}\bigl( g_c k - g_{\Lambda} u  \bigr) , 
\end{equation}
 leads to an Airy equation:
 \begin{equation}
\Biggl( \frac{\partial^2}{\partial \xi^2}  - \xi  \Biggr) \Psi(\xi) = 0, \label{AQWdW}
\end{equation}
whose solution is
\begin{equation}
\Psi(\xi) = C_1 Ai(\xi) + C_2 B(\xi) \, ,  \label{sol}
\end{equation}
where $Ai(\xi)$ and $B(\xi)$ are the Airy functions of the first and second kind, respectively.  
The system of equations also supports complex conjugated solutions:
\begin{equation}
\Psi^*(\xi^*) = C^*_1 Ai^*(\xi^*) + C^*_2 B^*(\xi^*) \, . \label{sol*} 
\end{equation}

\subsubsection{Boundary Conditions}
  
 The evolution phases of the universe can lead to different combinations of pressure and density characterised by different values of the dimensionless thermodynamical connection $\epsilon(t)$  (\ref{e}). In the following, however, for simplicity we chose a symmetric evolution description of the universe for positive and negative {\it cosmological time}. 

It is expected that the most appropriate solutions of the  WdW equation give rise, in the late universe, to a classic spacetime, and provide an initial condition for the inflationary period, necessary for the resolution of the flatness and horizon problems of classical cosmology. To meet these expectations, it is crucial to impose appropriate boundary conditions to the WdW equation. 
Here, as a first sketch of our quantum proposal, we do not enter in those specific 
aspects and impose the following boundary conditions:
\begin{equation}
\lim_{\xi \to -\infty}  \Psi(-\xi) \to 0 \, ; 
\lim_{\xi \to \infty} \Psi(\xi) \to 0 \, .
\end{equation}

\section{Results}

\begin{figure}[htbp]
\centering
\includegraphics[width=43mm,height=50mm]{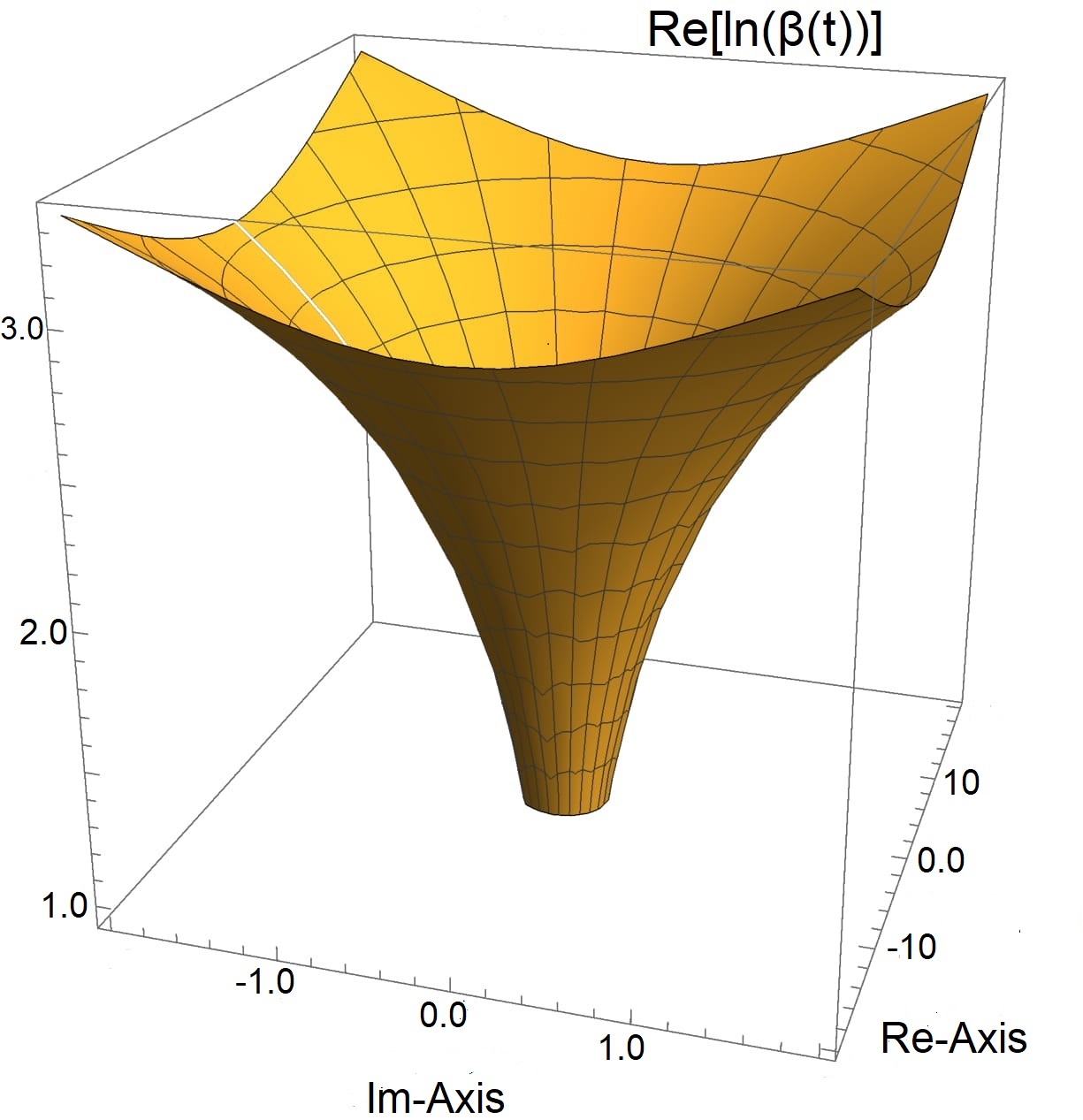}
\includegraphics[width=43mm,height=50mm]{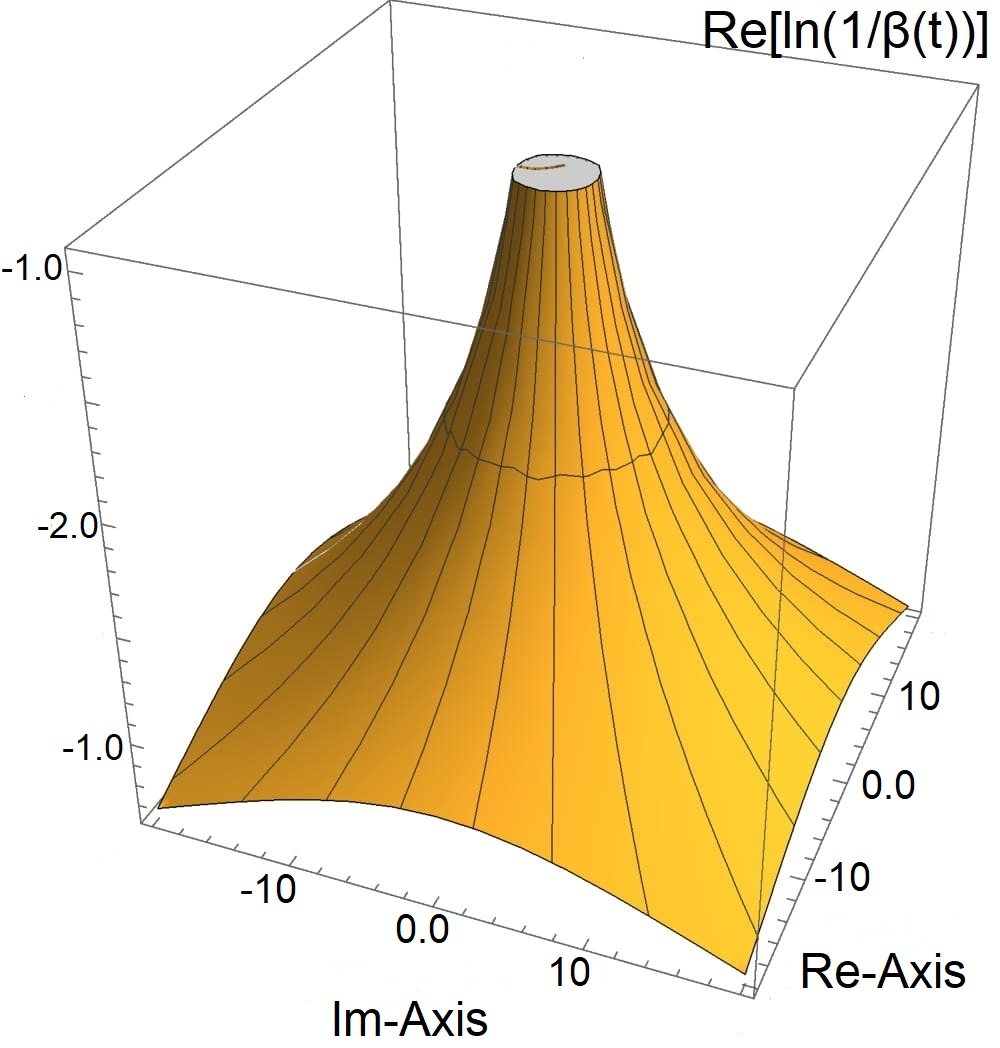}
\caption{Left figure: Characteristic plot of the Riemann surface $R$ associated to the real part of the $\ln[\beta(t)]$ function, represented by $Re[\ln[\beta(t)]]$, limited to one Riemann sheet in the region surrounding the Planck scale, the transition region corresponding to the domain where general relativity and quantum mechanics reconcile.  Right figure:  Plot of the real part of the $Re[\ln[1/\beta(t)]]$, assuming that the $\beta(t)$ function is orthomodular.  The domain of the quantum leap is indicated in the figure. 
} \label{fig1}
\end{figure}

\begin{figure}[htbp]
\centering
\includegraphics[width=43mm,height=45mm]{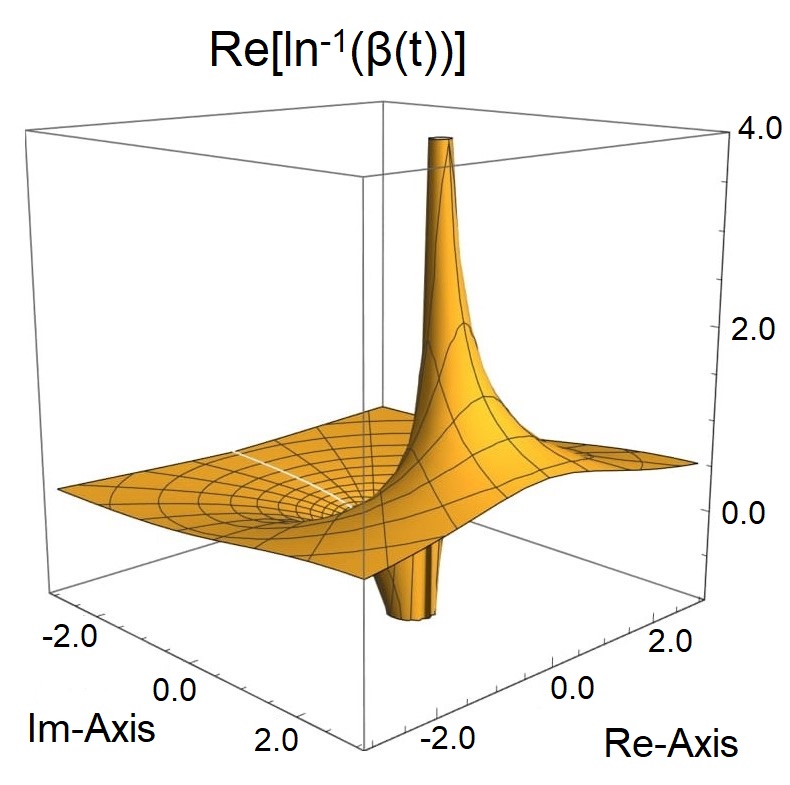}
\includegraphics[width=43mm,height=45mm]{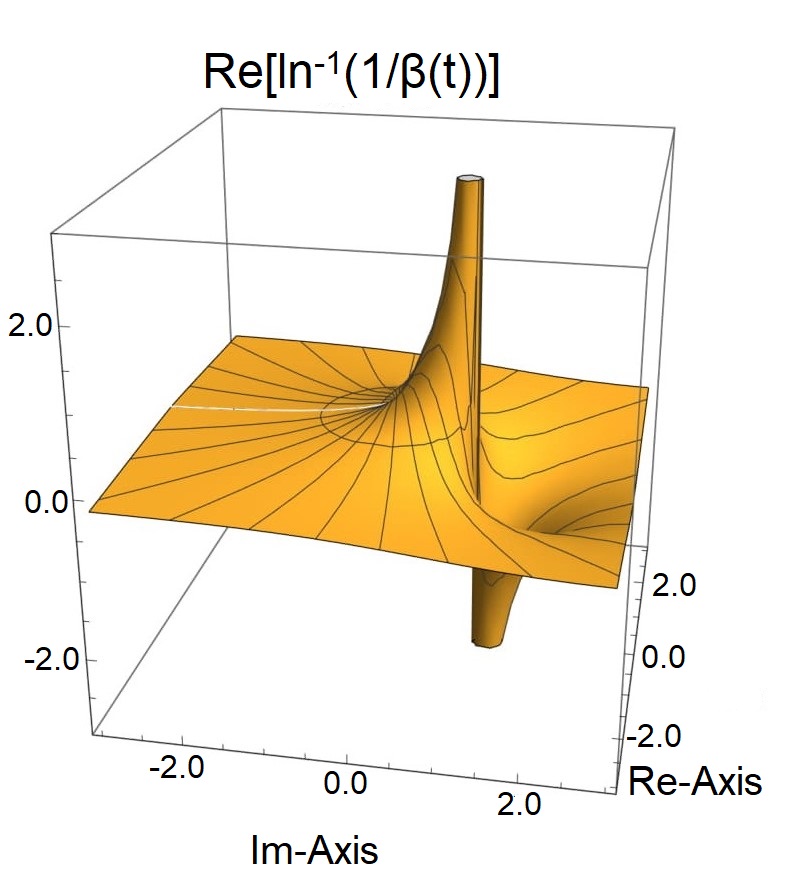}
\caption{Left figure: Characteristic plot of the Riemann surface $R$ associated to the real part of the $\ln^{-1}[\beta(t)]$ function, represented by $Re[\ln^{-1}[\beta(t)]]$ limited to one Riemann sheet in the region surrounding the Planck scale.   Right figure:  Plot of the real part of the inverse of the previous figure, $Re[\ln^{-1}[1/\beta(t)]]$.} \label{fig2}
\end{figure}

In Fig. (\ref{fig1}) we show characteristic plots of the Riemann surface associated to the real parts of $\ln[\beta(t)]$ and $\ln[1/\beta(t)]$, assuming that $\beta(t)$ is a orthomodular function. Fig. (\ref{fig2})shows the corresponding plots of $\ln^{-1}[\beta(t)]$ and $\ln^{-1}[1/\beta(t)]$. 

\begin{figure}[htb]
\centering
\includegraphics[width=42mm,height=42mm]{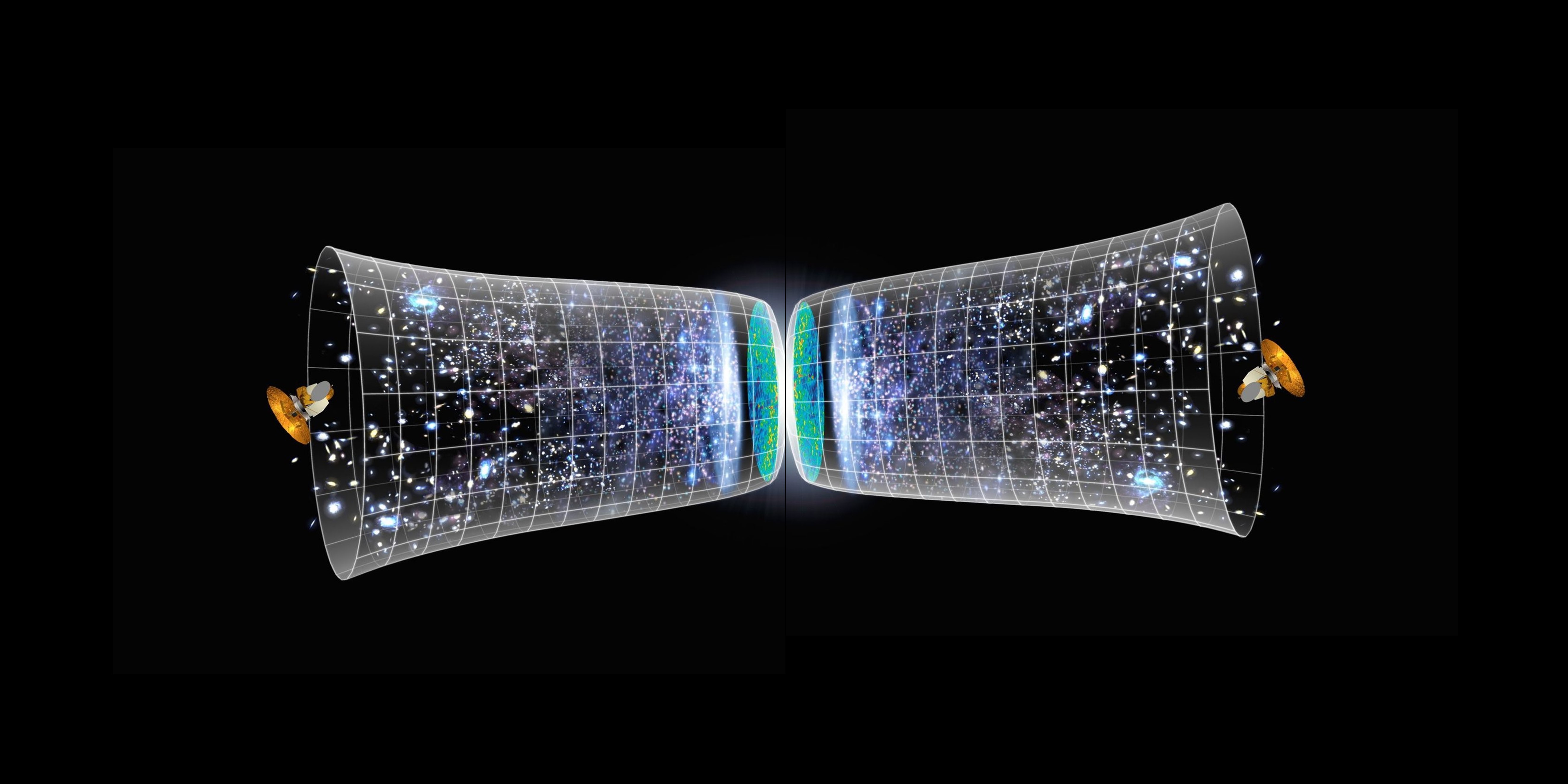}   \label{ESO1}
\includegraphics[width=42mm,height=42mm]{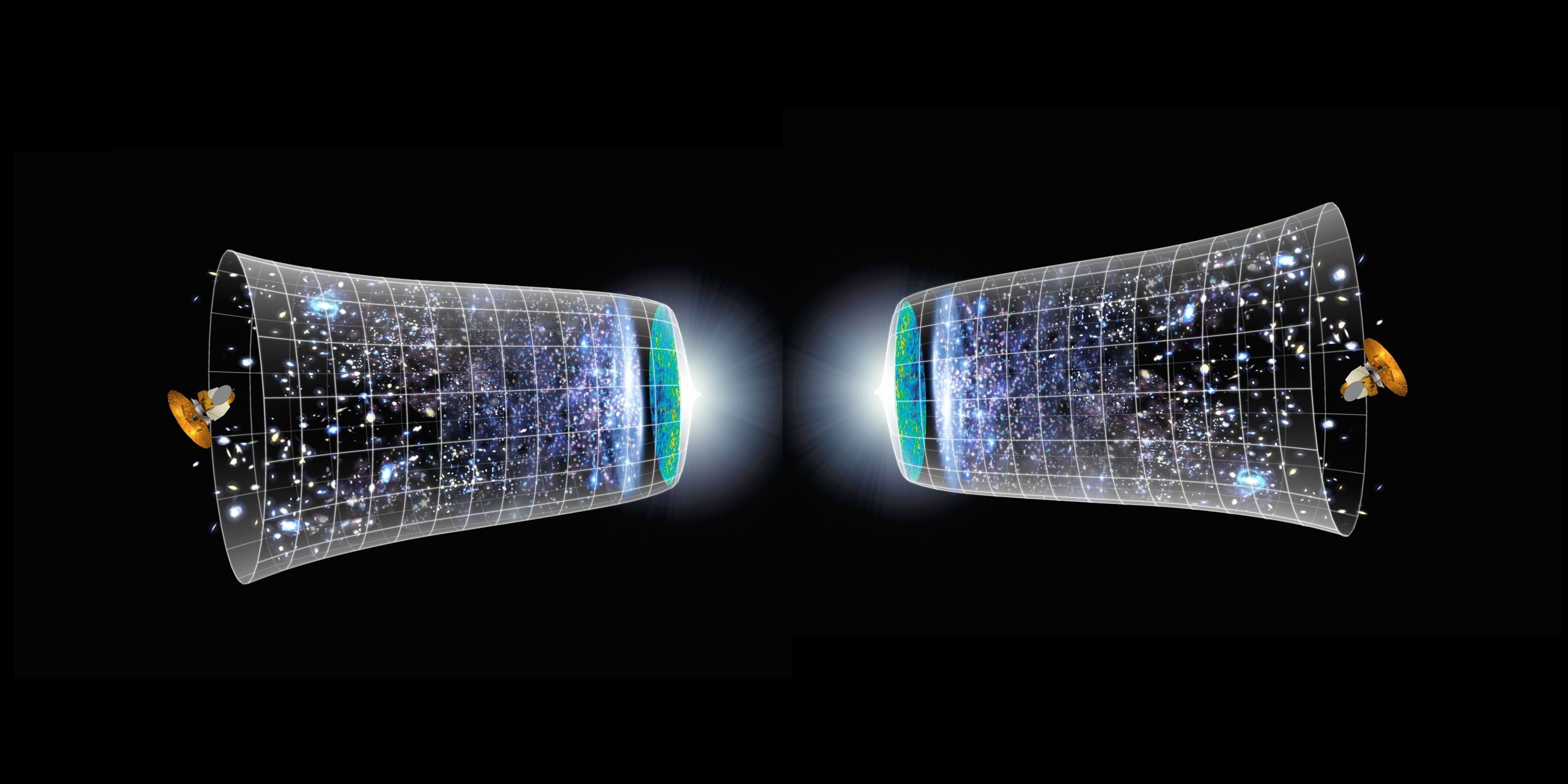}  \label{ESO2}
\caption{Left figure shows an artistic representation of the  {\it branch cut universe} evolution with two scenarios. 
The figures were based on the artistic impressions by ESO / M. Kornmesser~\citep{ESO}.} \label{ESO}
\end{figure}

\subsection{Scenarios for the branch cut universe} 

\begin{figure}[htbp]
\centering
\includegraphics[width=60mm,height=60mm]{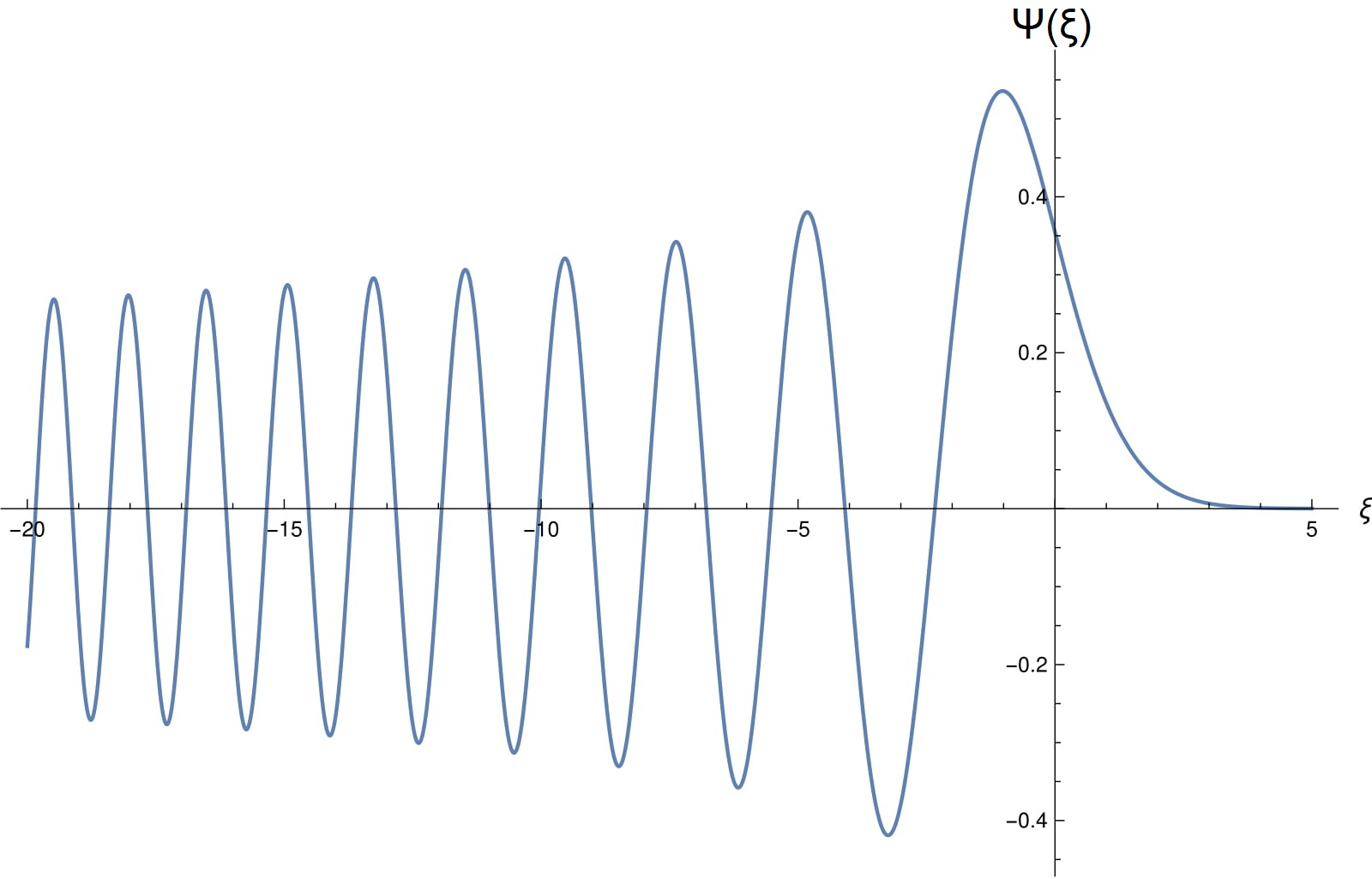}
\includegraphics[width=60mm,height=60mm]{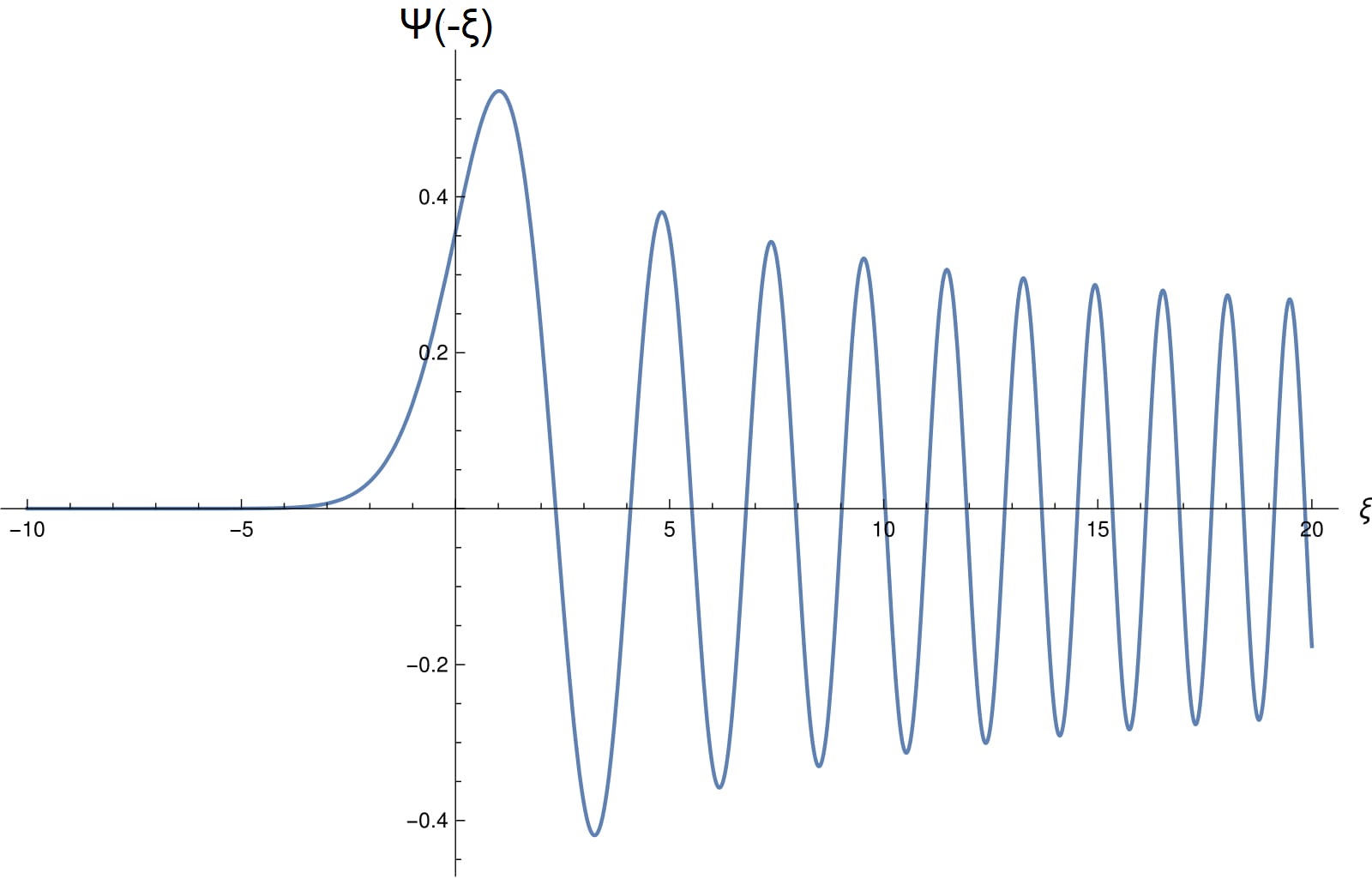}
\caption{Characteristic solution of equation (\ref{sol}).} \label{fig3}
\end{figure}

Our previous results delineate two scenarios for the evolution of the branch cut universe which are
sketched in an artistic representation 
(see Fig. (\ref{ESO})), with a branch point and a branch cut on the left figure, and no primordial singularity on the right one\footnote{Figures based on an artistic impression originally developed by ESO / M. Kornmesser~\citep{ESO}.}.
The figures indicate the cosmic contraction and expansion phases of the branch cut universe. In the representation sketched on the left figure, the branch cut universe evolves from negative to positive values of the complex cosmological time $t_C$, --- or the thermal time $T$, as a result of a Wick rotation ---, circumventing continuously a  a branch cut and no primordial singularity occurs, only branch points. The right figure on (\ref{ESO}) sketches an alternative artistic representation of two mirrored evolving universes, originated both from primordial singularities.  As an example,
in Fig. (\ref{fig3}) characteristic plots of the unnormalised solutions of equation (\ref{sol}), for the first scenario, are shown.

\section{Discussion}

The first scenario of the {\it branch cut universe} is characterised by its continuous expansion and by a systematic decrease of the temperature in its positive complex cosmological time sector. In the second scenario, the branch cut and branch point disappear after the {\it realisation} of imaginary time by means of a Wick rotation, which is replaced here by the real and continuous thermal time (temperature). In this second scenario, a mirrored parallel evolutionary universe, adjacent to ours, is nested in the structure of space and time, with its evolutionary process going backwards in the cosmological thermal time negative sector. In this case, the connection between the previous solutions is {\it broken} as a result of the Wick rotation. A similar result may be obtained if we adopt an approach based on the path integral formalism with no singularity in the first scenario. In the first scenario the entropy decreases systematically and continuously in the negative thermal time sector  until the absolute zero of entropy was reached. And then follows the increase of the entropy systematically in the positive thermal time sector. In the second scenario, entropy increases systematically in the evolution process of our universe but in the parallel mirror-universe, the arrow of time points down the entropy gradient, so the entropy is negative (negentropy sector).

Our sketched quantised formulation, based on the Wheeler De Witt equation, brings an alternative ingredient to overcoming the primordial singularity of the universe. 
In particular, the solutions of the quantised version (\ref{sol}) are in line with the description of the first scenario. A more detailed and in-depth discussion of these aspects involving both scenarios requires a more detailed future analysis.

\subsubsection{Observational signatures}

 An expressive challenge is the observational realisation of the proposal presented.  Speculations associated with the birth of two universes during the Big Bang, above 13.5 billion years ago, - our universe and another one, which from our perspective is functioning in reverse with time running backward ---,  as well as the multiverse conception are known and recurring. Fictional literature is lavish in this type of narrative, and from the scientific point of view, there are renowned scientists who are skeptical of the conception, and others who are proponents of multiverse theories, as S. Hawking for instance. 
  
Observations that may give some shelter to such conceptions are very rare or nonexistent. Interpretations of observational data based on such hypotheses were quickly demystified. More recently, the Antarctic Impulsive Transient Antenna ANITA/NASA project has detected for the second time~\citep{Gorham2018} a fountain of high-energy particles that resembles an upside-down cosmic-ray shower, generating a pleiades of speculations\footnote{The upward going cosmic rays, generated speculations running from sterile neutrinos and atypical dark matter distributions inside the Earth~\citep{SA} to a topsy-turvy universe created during the Big Bang and existing in parallel with ours~\citep{NS}.} about the meaning of these observations and the possible realisation of a universe specular to ours.  Although not supported by the authors of the article, speculations still persist\footnote{A subsequent article sought for a consistent explanation for the observed anomalies with no conclusive results up to now~\citep{Smith} .}. From our perspective, the formalism presented here represents a mathematical resource with a view to overcoming singularities in general relativity.
 
\section{Final remarks}
 
 As stressed before~\citep{ZenA}, the present formalism presents similarity with quantum bouncing models which assume in general a mechanism (or trigger) to keep the bouncing phase stable which could be associated for example with quantum fluctuations. In this kind of quantum bouncing models, the contraction phase amplifies quantum fluctuations and could serve as a trigger for the expansion phase (see for instance \citet{Novello2008}). 
  In turn, the quantum formulation sketched in this contribution represents a kind of quantum  tunnelling between the contraction and the expansion phases, an effect quite similar to the corresponding tunnelling effect in ordinary quantum mechanics.
  
 Using the causal structure of the McVittie spacetime~\citep{McVittie1933} for a classical bouncing cosmological model, \citet{Perez2021} recently found out that when the universe reaches a certain minimum scale, the trapping horizons disappear and the black hole ceases to exist, suggesting that neither a contracting nor an expanding universe can accommodate a black hole at all times. In our view, the formalism presented here could represent a solution for the survival of black holes from the contraction sector to the expanding one. The model presented by ~\citet{Perez2021} contains a fundamental ingredient for such description, the scale factor $a(t)$ that can be analytically continued to the complex space, thus enabling a transition from the contraction phase, in which the universe reaches a certain minimum scale that causes the trapping horizons to disappear and the black hole ceases to exist to the expansion phase. An investigation into this topi c is ongoing.  
 
 In physics, the prevailing tendency among scientists is to think of space and time as constituting the central structure of the universe. 
 Conceptions as physics without time and about the flow of time being an illusion have enriched the debate about its meaning. 
 A question then arises: how to reconcile these visions with the remarkable predictions of general relativity that implies a {\it materialisation} of spacetime, such as in the detection of gravitational waves, conceived as `ripples' in spacetime? We obviously do not intend to have a definitive answer to this question. As a final word, as we see, speculations on this still open questions find a fertile sea in another Einstein's quotation~\citep{Hedman2017}:
`time and space are modes by which we think and not conditions in which we live', a statement so powerful and profound that it will certainly continue to enlighten our creativity and imagination.    

\section{Acknowledgements}
P:O:H: acknowledges financial support from PAPIIT-DGAPA (IN100421).  The authors wish to thank the referees for valuable comments.
\appendix

\section{Cosmography parameters}\label{appendixB}

In the following we show our results for the analytically continued cosmography parameters for the radiation-, matter-, and dark-matter dominated eras.  

\subsection{Radiation-dominated era}

For the  radiation-dominated era, we obtain 
\begin{equation}
  \rho(t) \simeq  \frac{\rho_0}{\sqrt[2\epsilon]{\ln^{-2}[\beta(t_{\rm{P}})]   
+ \frac{1}{ \ln^{\, 2}(\beta_0)} \sqrt{\frac{2 \pi G \rho_0}{3}} \Bigl( t - t_{\rm{P}}  \Bigr) 
 } }. \label{rhoepsilon}
\end{equation}

Similarly,
\begin{equation}
  \rho^*(t^*) \simeq  \frac{\rho^*_0}{\sqrt[2\epsilon]{\ln^{-2}[\beta(t_{\rm{P}}^*)]   
+ \frac{1}{ \ln^{\, 2}(\beta_0)} \sqrt{\frac{2 \pi G \rho_0}{3}} \Bigl( t^* - t_{\rm{P}}^*  \Bigr)
 } }. 
\end{equation}

Additionally, we have
\begin{eqnarray}
& \Omega_{\rm{ccf}}(t) = -\frac{3k\ln^{\, 4}(\beta_0)}{ \pi G \rho_0} \frac{\ln^2\sqrt{  \ln^{-2}[\beta(t_{\rm{P}})]   
+ \frac{1}{ \ln^{\, 2}(\beta_0)} \sqrt{\frac{2 \pi G \rho_0}{3}} \Bigl( t - t_{\rm{P}} \Bigr) }}{\Bigl(\ln^{-2}[\beta(t_{\rm{P}})]   
+ \frac{1}{ \ln^{\, 2}(\beta_0)} \sqrt{\frac{2 \pi G \rho_0}{3}} \bigl( t - t_{\rm{P}}  \bigr)\Bigr)} \, , & \nonumber \\
\end{eqnarray}
and
\begin{eqnarray}
&
\Omega_{\rm{caf}}(t) =  \frac{3 \sigma^2\ln^{\, 4}(\beta_0)}{\pi G \rho_0} \frac{\ln^2\sqrt{  \ln^{-2}[\beta(t_{\rm{P}})]   
+ \frac{1}{ \ln^{\, 2}(\beta_0)} \sqrt{\frac{2 \pi G \rho_0}{3}} \Bigl( t - t_{\rm{P}} \Bigr) }}{\Bigl(\ln^{-2}[\beta(t_{\rm{P}})]   
+ \frac{1}{ \ln^{\, 2}(\beta_0)} \sqrt{\frac{2 \pi G \rho_0}{3}} \bigl( t - t_{\rm{P}}  \bigr)\Bigr)^4} \, . & \nonumber \\
\end{eqnarray}

Similarly, 
\begin{eqnarray}
&
\Omega^*_{\rm{ccf}}(t^*) = -\frac{3k\ln^{\, 4}(\beta^*_0)}{ \pi G \rho^*_0} \frac{\ln^2\sqrt{  \ln^{-2}[\beta(t^*_{\rm{P}})]   
+ \frac{1}{ \ln^{\, 2}(\beta^*_0)} \sqrt{\frac{2 \pi G \rho^*_0}{3}} \Bigl(t^* - t^*_{\rm{P}} \Bigr) }}{\Bigl(\ln^{-2}[\beta(t^*_{\rm{P}})]   
+ \frac{1}{ \ln^{\, 2}(\beta^*_0)} \sqrt{\frac{2 \pi G \rho^*_0}{3}} \bigl(t^* - t^*_{\rm{P}}  \bigr)\Bigr)} \!, & \nonumber \\
\end{eqnarray}
and
\begin{eqnarray}
&
\Omega^*_{\rm{caf}}(t^*) =  \frac{3 \sigma^2\ln^{\, 4}(\beta^*_0)}{\pi G \rho^*_0} \frac{\ln^2\sqrt{  \ln^{-2}[\beta(t^*_{\rm{P}})]   
+ \frac{1}{ \ln^{\, 2}(\beta^*_0)} \sqrt{\frac{2 \pi G \rho^*_0}{3}} \Bigl( t^* - t^*_{\rm{P}} \Bigr) }}{\Bigl(\ln^{-2}[\beta(t^*_{\rm{P}})]   
+ \frac{1}{ \ln^{\, 2}(\beta^*_0)} \sqrt{\frac{2 \pi G \rho^*_0}{3}} \bigl( t^* - t^*_{\rm{P}}  \bigr)\Bigr)^4} . & \nonumber \\
\end{eqnarray}

\subsection{Matter-dominated era}

For the  matter-dominated era, the 
following result holds
  (\ref{dens}) give
 \begin{equation}
  \rho(t) \simeq    \frac{\rho^*_0}{\sqrt[4\epsilon/3]{ \ln^{-3/2}[\beta(t_{\rm{P}})]   
 +  \frac{1}{ \ln^{\, 3/2}(\beta_0)}  \sqrt{6 \pi G \rho_0} \Bigl(t - t_{\rm{P}} \Bigr)
 } } \, . 
\end{equation}
Similarly
 \begin{equation}
  \rho^*(t^*) \simeq    \frac{\rho_0}{\sqrt[4\epsilon/3]{ \ln^{-3/2}[\beta(t_{\rm{P}}^*)]   
 +  \frac{1}{ \ln^{\, 3/2}(\beta_0)}  \sqrt{6 \pi G \rho_0} \Bigl(t^* - t_{\rm{P}}^* \Bigr)
 } } \, .
\end{equation}
Additionally we obtain
 \begin{eqnarray}
& \Omega_{\rm{ccf}}(t) 
  = - \frac{k\ln^3(\beta_0)}{4 \pi G \rho_0} \frac{ \ln^2\Biggl[  \sqrt[2/3]{ \ln^{-3/2}[\beta(t_{\rm{P}})]   
 +  \frac{1}{ \ln^{\, 3/2}(\beta_0)}  \sqrt{6 \pi G \rho_0} \Bigl(t - t_{\rm{P}} \Bigr)
 }  \Biggr] }{\sqrt[4/3]{ \ln^{-3/2}[\beta(t_{\rm{P}})]   
 +  \frac{1}{ \ln^{\, 3/2}(\beta_0)}  \sqrt{6 \pi G \rho_0} \Bigl(t - t_{\rm{P}} \Bigr)}},  & \nonumber \\
\end{eqnarray}
and
 \begin{eqnarray}
& \Omega_{\rm{caf}}(t) = \frac{\sigma^2\ln^3(\beta_0)}{4 \pi G \rho_0} \frac{ \ln^2\Biggl[  \sqrt[2/3]{ \ln^{-3/2}[\beta(t_{\rm{P}})]   
 +  \frac{1}{ \ln^{\, 3/2}(\beta_0)}  \sqrt{6 \pi G \rho_0} \Bigl(t - t_{\rm{P}} \Bigr)
 }  \Biggr] }{\sqrt[8/3]{ \ln^{-3/2}[\beta(t_{\rm{P}})]   
 +  \frac{1}{ \ln^{\, 3/2}(\beta_0)}  \sqrt{6 \pi G \rho_0} \Bigl(t - t_{\rm{P}} \Bigr)}},  & \nonumber \\
\end{eqnarray}
Similarly we get
 \begin{eqnarray}
 &
\Omega^*_{\rm{ccf}}(t^*) \! = \! - \frac{k\ln^3(\beta^*_0)}{4 \pi G \rho^*_0} \frac{ \ln^2\Biggl[  \sqrt[2/3]{ \ln^{-3/2}[\beta^*(t^*_{\rm{P}})]   
 +  \frac{1}{ \ln^{\, 3/2}(\beta^*_0)}  \sqrt{6 \pi G \rho^*_0} \Bigl(t^* - t^*_{\rm{P}} \Bigr)
 }  \Biggr] }{\sqrt[4/3]{ \ln^{-3/2}[\beta^*(t^*_{\rm{P}})]   
 +  \frac{1}{ \ln^{\, 3/2}(\beta^*_0)}  \sqrt{6 \pi G \rho^*_0} \Bigl(t^* - t^*_{\rm{P}} \Bigr)}},  & \nonumber \\
\end{eqnarray}
and
 \begin{eqnarray}
& \Omega^*_{\rm{caf}}(t^*) =  \frac{\sigma^2\ln^3(\beta^*_0)}{4 \pi G \rho^*_0} \frac{ \ln^2\Biggl[  \sqrt[2/3]{ \ln^{-3/2}[\beta^*(t^*_{\rm{P}})]   
 +  \frac{1}{ \ln^{\, 3/2}(\beta^*_0)}  \sqrt{6 \pi G \rho^*_0} \Bigl(t^* - t^*_{\rm{P}} \Bigr)
 }  \Biggr] }{\sqrt[8/3]{ \ln^{-3/2}[\beta^*(t^*_{\rm{P}})]   
 +  \frac{1}{ \ln^{\, 3/2}[\beta^*_0)}  \sqrt{6 \pi G \rho^*_0} \Bigl(t^* - t^*_{\rm{P}} \Bigr)}},  & \nonumber \\
\end{eqnarray}

\bibliography{ZenB}%

\begin{thebibliography}{}

\bibitem [\protect \citeauthoryear {%
Bertolami%
\ \BBA {} Zarro%
}{%
Bertolami%
\ \BBA {} Zarro%
}{%
{\protect \APACyear {2011}}%
}]{%
Orfeu}
\APACinsertmetastar {%
Orfeu}%
\begin{APACrefauthors}%
Bertolami, O.%
\BCBT {}\ \BBA {} Zarro, C\BPBI A\BPBI D.%
\end{APACrefauthors}%
\unskip\
\newblock
\APACrefYearMonthDay{2011}{}{},
\newblock
\unskip
\newblock
\APACjournalVolNumPages{Phys. Rev. D}{84}{}{044042}.
\PrintBackRefs{\CurrentBib}

\bibitem [\protect \citeauthoryear {%
Calmet%
, Graesser%
\BCBL {}\ \BBA {} Hsu%
}{%
Calmet%
\ \protect \BOthers {.}}{%
{\protect \APACyear {(2004}}%
}]{%
Calmet}
\APACinsertmetastar {%
Calmet}%
\begin{APACrefauthors}%
Calmet, X.%
, Graesser, M.%
\BCBL {}\ \BBA {} Hsu, S\BPBI D\BPBI H.%
\end{APACrefauthors}%
\unskip\
\newblock
\APACrefYearMonthDay{(2004}{}{},
\newblock
\unskip
\newblock
\APACjournalVolNumPages{Phys. Rev. Lett.}{93}{}{211101}.
\PrintBackRefs{\CurrentBib}

\bibitem [\protect \citeauthoryear {%
Cartwright%
}{%
Cartwright%
}{%
{\protect \APACyear {2020}}%
}]{%
NS}
\APACinsertmetastar {%
NS}%
\begin{APACrefauthors}%
Cartwright, J.%
\end{APACrefauthors}%
\unskip\
\newblock
\APACrefYearMonthDay{2020}{}{},
\newblock
\APACrefbtitle {We may have spotted a parallel universe going backwards in
  time.} {We may have spotted a parallel universe going backwards in time.},
\newblock
\APAChowpublished {Available at
  \url{https://www.newscientist.com/article/mg24532770-400-we-may-have-spotted-a-parallel-universe-going-backwards-in-time/}
  (New Scientist)}.
\PrintBackRefs{\CurrentBib}

\bibitem [\protect \citeauthoryear {%
Cordero%
, Garcia-Compean%
\BCBL {}\ \BBA {} Turrubiates%
}{%
Cordero%
\ \protect \BOthers {.}}{%
{\protect \APACyear {2019}}%
}]{%
Cordero}
\APACinsertmetastar {%
Cordero}%
\begin{APACrefauthors}%
Cordero, R.%
, Garcia-Compean, H.%
\BCBL {}\ \BBA {} Turrubiates, F\BPBI J.%
\end{APACrefauthors}%
\unskip\
\newblock
\APACrefYearMonthDay{2019}{}{},
\newblock
\unskip
\newblock
\APACjournalVolNumPages{General Relativity and Gravitation}{51}{}{138}.
\PrintBackRefs{\CurrentBib}

\bibitem [\protect \citeauthoryear {%
DeWitt%
}{%
DeWitt%
}{%
{\protect \APACyear {1967}}%
}]{%
WdW}
\APACinsertmetastar {%
WdW}%
\begin{APACrefauthors}%
DeWitt, B\BPBI S.%
\end{APACrefauthors}%
\unskip\
\newblock
\APACrefYearMonthDay{1967}{}{},
\newblock
\unskip
\newblock
\APACjournalVolNumPages{Phys. Rev.}{160}{}{1113}.
\PrintBackRefs{\CurrentBib}

\bibitem [\protect \citeauthoryear {%
Einstein%
}{%
Einstein%
}{%
{\protect \APACyear {2020}}%
}]{%
CHRISTIES2020}
\APACinsertmetastar {%
CHRISTIES2020}%
\begin{APACrefauthors}%
Einstein, A.%
\end{APACrefauthors}%
\unskip\
\newblock
\APACrefYear{2020},
\newblock
\APACrefbtitle {Time's arrow: Albert Einstein's letters to Michele Besso}
  {Time's arrow: Albert Einstein's letters to Michele Besso}.
\newblock
\APACaddressPublisher{Los Angeles, USA}{Christie's}.
\PrintBackRefs{\CurrentBib}

\bibitem [\protect \citeauthoryear {%
Feinberg%
\ \BBA {} Peleg%
}{%
Feinberg%
\ \BBA {} Peleg%
}{%
{\protect \APACyear {1995}}%
}]{%
Feinberg}
\APACinsertmetastar {%
Feinberg}%
\begin{APACrefauthors}%
Feinberg, J.%
\BCBT {}\ \BBA {} Peleg, Y.%
\end{APACrefauthors}%
\unskip\
\newblock
\APACrefYearMonthDay{1995}{}{},
\newblock
\unskip
\newblock
\APACjournalVolNumPages{Phys.Rev. D}{52}{}{1988}.
\PrintBackRefs{\CurrentBib}

\bibitem [\protect \citeauthoryear {%
Feynman%
\ \BBA {} Hibbs%
}{%
Feynman%
\ \BBA {} Hibbs%
}{%
{\protect \APACyear {1965}}%
}]{%
Feynman1965}
\APACinsertmetastar {%
Feynman1965}%
\begin{APACrefauthors}%
Feynman, R.%
\BCBT {}\ \BBA {} Hibbs, A.%
\end{APACrefauthors}%
\unskip\
\newblock
\APACrefYear{1965},
\newblock
\APACrefbtitle {Quantum Mechanics and Path Integrals} {Quantum Mechanics and
  Path Integrals}.
\newblock
\APACaddressPublisher{New York, USA}{McGraw-Hill}.
\PrintBackRefs{\CurrentBib}

\bibitem [\protect \citeauthoryear {%
Garay%
}{%
Garay%
}{%
{\protect \APACyear {1999}}%
}]{%
Garay1999}
\APACinsertmetastar {%
Garay1999}%
\begin{APACrefauthors}%
Garay, L\BPBI J.%
\end{APACrefauthors}%
\unskip\
\newblock
\APACrefYearMonthDay{1999}{}{},
\newblock
\unskip
\newblock
\APACjournalVolNumPages{Int. J. Mod. Phys. A}{14}{}{4079}.
\PrintBackRefs{\CurrentBib}

\bibitem [\protect \citeauthoryear {%
Gorham%
\ \protect \BOthers {.}}{%
Gorham%
\ \protect \BOthers {.}}{%
{\protect \APACyear {2018}}%
}]{%
Gorham2018}
\APACinsertmetastar {%
Gorham2018}%
\begin{APACrefauthors}%
Gorham, P\BPBI W.%
, Rotter, B.%
, Allison, P.%
\ et al.\end{APACrefauthors}%
\unskip\
\newblock
\APACrefYearMonthDay{2018}{}{},
\newblock
\unskip
\newblock
\APACjournalVolNumPages{Phys. Rev. Lett.}{121}{}{161102}.
\PrintBackRefs{\CurrentBib}

\bibitem [\protect \citeauthoryear {%
He%
\ \BBA {} Cai%
}{%
He%
\ \BBA {} Cai%
}{%
{\protect \APACyear {2020}}%
}]{%
He}
\APACinsertmetastar {%
He}%
\begin{APACrefauthors}%
He, D.%
\BCBT {}\ \BBA {} Cai, Q.%
\end{APACrefauthors}%
\unskip\
\newblock
\APACrefYearMonthDay{2020}{}{},
\newblock
\unskip
\newblock
\APACjournalVolNumPages{Physics Letters B}{809}{}{135747}.
\PrintBackRefs{\CurrentBib}

\bibitem [\protect \citeauthoryear {%
Hedman%
}{%
Hedman%
}{%
{\protect \APACyear {2017}}%
}]{%
Hedman2017}
\APACinsertmetastar {%
Hedman2017}%
\begin{APACrefauthors}%
Hedman, A.%
\end{APACrefauthors}%
\unskip\
\newblock
\APACrefYear{2017},
\newblock
\APACrefbtitle {Consciousness from a Broad Perspective} {Consciousness from a
  Broad Perspective}.
\newblock
\APACaddressPublisher{Berlin, Germany}{Springer}.
\PrintBackRefs{\CurrentBib}

\bibitem [\protect \citeauthoryear {%
Hess%
}{%
Hess%
}{%
{\protect \APACyear {2017}}%
}]{%
Hess2017}
\APACinsertmetastar {%
Hess2017}%
\begin{APACrefauthors}%
Hess, P\BPBI O.%
\end{APACrefauthors}%
\unskip\
\newblock
\APACrefYear{2017},
\newblock
\APACrefbtitle {Centennial of General Relativity: A Celebration} {Centennial of
  General Relativity: A Celebration}\ (C\BPBI A\BPBI Z.~Vasconcellos, \BED{}).
\newblock
\APACaddressPublisher{Singapore}{World Scientific Pub. Co.}
\PrintBackRefs{\CurrentBib}

\bibitem [\protect \citeauthoryear {%
Hess%
\ \BBA {} Boller%
}{%
Hess%
\ \BBA {} Boller%
}{%
{\protect \APACyear {2020}}%
}]{%
Hess2020}
\APACinsertmetastar {%
Hess2020}%
\begin{APACrefauthors}%
Hess, P\BPBI O.%
\BCBT {}\ \BBA {} Boller, T.%
\end{APACrefauthors}%
\unskip\
\newblock
\APACrefYear{2020},
\newblock
\APACrefbtitle {Topics on Strong Gravity: A Modern View on Theories and
  Experiments} {Topics on Strong Gravity: A Modern View on Theories and
  Experiments}\ (C\BPBI A\BPBI Z.~Vasconcellos, \BED{}).
\newblock
\APACaddressPublisher{Singapore}{World Scientific Pub. Co.}
\PrintBackRefs{\CurrentBib}

\bibitem [\protect \citeauthoryear {%
Hess%
, Sch\"afer%
\BCBL {}\ \BBA {} Greiner%
}{%
Hess%
\ \protect \BOthers {.}}{%
{\protect \APACyear {2015}}%
}]{%
PNP2020}
\APACinsertmetastar {%
PNP2020}%
\begin{APACrefauthors}%
Hess, P\BPBI O.%
, Sch\"afer, M.%
\BCBL {}\ \BBA {} Greiner, W.%
\end{APACrefauthors}%
\unskip\
\newblock
\APACrefYear{2015},
\newblock
\APACrefbtitle {Pseudo-Complex General Relativity} {Pseudo-Complex General
  Relativity}.
\newblock
\APACaddressPublisher{Heidelberg, Berlin, Germany}{Springer}.
\PrintBackRefs{\CurrentBib}

\bibitem [\protect \citeauthoryear {%
Isham%
}{%
Isham%
}{%
{\protect \APACyear {1993}}%
}]{%
Isham1993}
\APACinsertmetastar {%
Isham1993}%
\begin{APACrefauthors}%
Isham, C\BPBI J.%
\end{APACrefauthors}%
\unskip\
\newblock
\APACrefYear{1993},
\newblock
\APACrefbtitle {Canonical Quantum Gravity and the Problem of Time} {Canonical
  Quantum Gravity and the Problem of Time}\ (L\BPBI A.~Ibort\ \BBA {} M\BPBI
  A.~Rodríguez, \BEDS{}).
\newblock
\APACaddressPublisher{Netherlands}{Dordrecht: Springer}.
\PrintBackRefs{\CurrentBib}

\bibitem [\protect \citeauthoryear {%
Kim%
}{%
Kim%
}{%
{\protect \APACyear {1997}}%
}]{%
Kim}
\APACinsertmetastar {%
Kim}%
\begin{APACrefauthors}%
Kim, S\BPBI P.%
\end{APACrefauthors}%
\unskip\
\newblock
\APACrefYearMonthDay{1997}{}{},
\newblock
\unskip
\newblock
\APACjournalVolNumPages{Phys. Lett. A}{236}{}{11}.
\PrintBackRefs{\CurrentBib}

\bibitem [\protect \citeauthoryear {%
Kornmesser%
}{%
Kornmesser%
}{%
{\protect \APACyear {2020}}%
}]{%
ESO}
\APACinsertmetastar {%
ESO}%
\begin{APACrefauthors}%
Kornmesser, M.%
\end{APACrefauthors}%
\unskip\
\newblock
\APACrefYearMonthDay{2020}{}{},
\newblock
\APACrefbtitle {History of the Universe (ESO).} {History of the Universe
  (ESO).},
\newblock
\APAChowpublished {Available at
  \url{https://supernova.eso.org/exhibition/1101/}}.
\PrintBackRefs{\CurrentBib}

\bibitem [\protect \citeauthoryear {%
Letzter%
}{%
Letzter%
}{%
{\protect \APACyear {2018}}%
}]{%
SA}
\APACinsertmetastar {%
SA}%
\begin{APACrefauthors}%
Letzter, R.%
\end{APACrefauthors}%
\unskip\
\newblock
\APACrefYearMonthDay{2018}{}{},
\newblock
\APACrefbtitle {Bizarre particles keep flying out of Antarctica's ice, and they
  might shatter modern physics.} {Bizarre particles keep flying out of
  Antarctica's ice, and they might shatter modern physics.},
\newblock
\APAChowpublished {Available at
  \url{https://www.scientificamerican.com/article/bizarre-particles-keep-flying-out-of-antarcticas-ice-and-they-might-shatter-modern-physics/}
  (Scientific American)}.
\PrintBackRefs{\CurrentBib}

\bibitem [\protect \citeauthoryear {%
McVittie%
}{%
McVittie%
}{%
{\protect \APACyear {1933}}%
}]{%
McVittie1933}
\APACinsertmetastar {%
McVittie1933}%
\begin{APACrefauthors}%
McVittie, G.%
\end{APACrefauthors}%
\unskip\
\newblock
\APACrefYearMonthDay{1933}{}{},
\newblock
\unskip
\newblock
\APACjournalVolNumPages{Mon. Not. R. Astron. Soc.}{93}{}{}.
\PrintBackRefs{\CurrentBib}

\bibitem [\protect \citeauthoryear {%
Minkowski%
}{%
Minkowski%
}{%
{\protect \APACyear {1915}}%
}]{%
Minkowski1915}
\APACinsertmetastar {%
Minkowski1915}%
\begin{APACrefauthors}%
Minkowski, H.%
\end{APACrefauthors}%
\unskip\
\newblock
\APACrefYearMonthDay{1915}{}{},
\newblock
\unskip
\newblock
\APACjournalVolNumPages{Annalen der Physic}{47}{}{927}.
\PrintBackRefs{\CurrentBib}

\bibitem [\protect \citeauthoryear {%
Novello%
\ \BBA {} Bergliaffa%
}{%
Novello%
\ \BBA {} Bergliaffa%
}{%
{\protect \APACyear {2008}}%
}]{%
Novello2008}
\APACinsertmetastar {%
Novello2008}%
\begin{APACrefauthors}%
Novello, M.%
\BCBT {}\ \BBA {} Bergliaffa, S.%
\end{APACrefauthors}%
\unskip\
\newblock
\APACrefYearMonthDay{2008}{}{},
\newblock
\unskip
\newblock
\APACjournalVolNumPages{Physics Reports}{463 (4)}{}{127}.
\PrintBackRefs{\CurrentBib}

\bibitem [\protect \citeauthoryear {%
P\'erez%
, Bergliaffa%
\BCBL {}\ \BBA {} Romero%
}{%
P\'erez%
\ \protect \BOthers {.}}{%
{\protect \APACyear {2021}}%
}]{%
Perez2021}
\APACinsertmetastar {%
Perez2021}%
\begin{APACrefauthors}%
P\'erez, D.%
, Bergliaffa, S\BPBI E\BPBI P.%
\BCBL {}\ \BBA {} Romero, G\BPBI E.%
\end{APACrefauthors}%
\unskip\
\newblock
\APACrefYearMonthDay{2021}{}{},
\newblock
\unskip
\newblock
\APACjournalVolNumPages{Phys. Rev. D}{103}{}{064019}.
\PrintBackRefs{\CurrentBib}

\bibitem [\protect \citeauthoryear {%
Rovelli%
}{%
Rovelli%
}{%
{\protect \APACyear {2004}}%
}]{%
Rovelli2004}
\APACinsertmetastar {%
Rovelli2004}%
\begin{APACrefauthors}%
Rovelli, C.%
\end{APACrefauthors}%
\unskip\
\newblock
\APACrefYear{2004},
\newblock
\APACrefbtitle {Quantum Gravity} {Quantum Gravity}.
\newblock
\APACaddressPublisher{Cambridge, UK}{Cambridge University Press}.
\PrintBackRefs{\CurrentBib}

\bibitem [\protect \citeauthoryear {%
Rovelli%
}{%
Rovelli%
}{%
{\protect \APACyear {2015}}%
}]{%
Rovelli2015}
\APACinsertmetastar {%
Rovelli2015}%
\begin{APACrefauthors}%
Rovelli, C.%
\end{APACrefauthors}%
\unskip\
\newblock
\APACrefYearMonthDay{2015}{}{},
\newblock
\unskip
\newblock
\APACjournalVolNumPages{Classical and Quantum Gravity}{32}{}{124005}.
\PrintBackRefs{\CurrentBib}

\bibitem [\protect \citeauthoryear {%
Rovelli%
}{%
Rovelli%
}{%
{\protect \APACyear {2019}}%
}]{%
Rovelli2019}
\APACinsertmetastar {%
Rovelli2019}%
\begin{APACrefauthors}%
Rovelli, C.%
\end{APACrefauthors}%
\unskip\
\newblock
\APACrefYear{2019},
\newblock
\APACrefbtitle {The Order of Time} {The Order of Time}.
\newblock
\APACaddressPublisher{New York, USA}{Riverhead Books}.
\PrintBackRefs{\CurrentBib}

\bibitem [\protect \citeauthoryear {%
Rovelli%
\ \BBA {} Smerlak%
}{%
Rovelli%
\ \BBA {} Smerlak%
}{%
{\protect \APACyear {2011}}%
}]{%
Rovelli2011}
\APACinsertmetastar {%
Rovelli2011}%
\begin{APACrefauthors}%
Rovelli, C.%
\BCBT {}\ \BBA {} Smerlak, M.%
\end{APACrefauthors}%
\unskip\
\newblock
\APACrefYearMonthDay{2011}{}{},
\newblock
\unskip
\newblock
\APACjournalVolNumPages{Classical and Quantum Gravity}{28}{}{075007}.
\PrintBackRefs{\CurrentBib}

\bibitem [\protect \citeauthoryear {%
Sch\"onenbach%
\ \protect \BOthers {.}}{%
Sch\"onenbach%
\ \protect \BOthers {.}}{%
{\protect \APACyear {2014}}%
}]{%
MNRAS2014}
\APACinsertmetastar {%
MNRAS2014}%
\begin{APACrefauthors}%
Sch\"onenbach, T.%
, Caspar, G.%
, Hess, P\BPBI O.%
, Boller, T.%
, M\"uller, A.%
, Sch\"afer, M.%
\BCBL {}\ \BBA {} Greiner, W.%
\end{APACrefauthors}%
\unskip\
\newblock
\APACrefYearMonthDay{2014}{}{},
\newblock
\unskip
\newblock
\APACjournalVolNumPages{Month. Not. of the Roy. Astron. Society}{442}{}{}.
\PrintBackRefs{\CurrentBib}

\bibitem [\protect \citeauthoryear {%
Shestakova%
}{%
Shestakova%
}{%
{\protect \APACyear {2018}}%
}]{%
Shestakova2018}
\APACinsertmetastar {%
Shestakova2018}%
\begin{APACrefauthors}%
Shestakova, T\BPBI P.%
\end{APACrefauthors}%
\unskip\
\newblock
\APACrefYearMonthDay{2018}{}{},
\newblock
\unskip
\newblock
\APACjournalVolNumPages{Int. J. Mod. Phys. D}{27}{}{1841004}.
\PrintBackRefs{\CurrentBib}

\bibitem [\protect \citeauthoryear {%
Smith%
\ \protect \BOthers {.}}{%
Smith%
\ \protect \BOthers {.}}{%
{\protect \APACyear {2020}}%
}]{%
Smith}
\APACinsertmetastar {%
Smith}%
\begin{APACrefauthors}%
Smith, D.%
, Besson, D\BPBI Z.%
, Deaconu, C.%
\ et al.\end{APACrefauthors}%
\unskip\
\newblock
\APACrefYearMonthDay{2020}{9}{},
\newblock
\APACrefbtitle {{Experimental tests of sub-surface reflectors as an explanation
  for the ANITA anomalous events}.} {{Experimental tests of sub-surface
  reflectors as an explanation for the ANITA anomalous events}.}
\newblock
\APACrefnote{eprint 2009.13010, arXiv astro-ph.HE}
\PrintBackRefs{\CurrentBib}

\bibitem [\protect \citeauthoryear {%
Vasconcellos%
, Hadjimichef%
, Razeira%
, Volkmer%
\BCBL {}\ \BBA {} Bodmann%
}{%
Vasconcellos%
\ \protect \BOthers {.}}{%
{\protect \APACyear {2020}}%
}]{%
Zen2020}
\APACinsertmetastar {%
Zen2020}%
\begin{APACrefauthors}%
Vasconcellos, C\BPBI Z.%
, Hadjimichef, D.%
, Razeira, M.%
, Volkmer, G.%
\BCBL {}\ \BBA {} Bodmann, B.%
\end{APACrefauthors}%
\unskip\
\newblock
\APACrefYearMonthDay{2020}{}{},
\newblock
\unskip
\newblock
\APACjournalVolNumPages{Astronomische Nachrichten}{340 (9,10)}{}{857}.
\PrintBackRefs{\CurrentBib}

\bibitem [\protect \citeauthoryear {%
Vasconcellos%
\ \protect \BOthers {.}}{%
Vasconcellos%
\ \protect \BOthers {.}}{%
{\protect \APACyear {2021}}%
}]{%
ZenA}
\APACinsertmetastar {%
ZenA}%
\begin{APACrefauthors}%
Vasconcellos, C\BPBI Z.%
, Hess, P.%
, Hadjimichef, D.%
, Bodmann, B.%
, Razeira, M.%
\BCBL {}\ \BBA {} Volkmer, G\BPBI L.%
\end{APACrefauthors}%
\unskip\
\newblock
\APACrefYearMonthDay{2021}{}{},
\newblock
\unskip
\newblock
\APACjournalVolNumPages{Astronomische Nachrichten}{}{}{}.
\newblock
\APACrefnote{Accepted for publication}
\PrintBackRefs{\CurrentBib}

\end{thebibliography}

\end{document}